%% file: main.tex
\documentclass[conference]{IEEEtran}
\IEEEoverridecommandlockouts
\newcommand{\Epsilon}{\mathcal{E}}
\usepackage{cite}
\usepackage{tabularx}
\usepackage{graphicx}
\usepackage{mathtools}
\usepackage{amsmath,amssymb,amsfonts}
\usepackage{algorithm2e}
\usepackage{wasysym}
\usepackage{tablefootnote}
\usepackage{multicol}
\usepackage{caption}
\usepackage{subcaption}
\usepackage{textcomp}
\usepackage{import}
\usepackage{xcolor}
\usepackage{etoolbox}
\usepackage{listings}
\usepackage{hyperref}
\usepackage{booktabs}
\usepackage[capitalise,noabbrev]{cleveref}    
\usepackage{pgfplotstable}
\usepackage{lipsum}
\usepackage{balance}
\usepackage{import}
\usepackage[inline]{enumitem}
\usepackage{titlesec}
\usepackage{chapterfolder}
\usepackage{comment}
\usepackage{enumitem}
\usepackage{tikz}
\usetikzlibrary{patterns}

\definecolor[named]{ACMBlue}{cmyk}{1,0.1,0,0.1}
\definecolor[named]{ACMYellow}{cmyk}{0,0.16,1,0}
\definecolor[named]{ACMOrange}{cmyk}{0,0.42,1,0.01}
\definecolor[named]{ACMRed}{cmyk}{0,0.90,0.86,0}
\definecolor[named]{ACMLightBlue}{cmyk}{0.49,0.01,0,0}
\definecolor[named]{ACMGreen}{cmyk}{0.20,0,1,0.19}
\definecolor[named]{ACMPurple}{cmyk}{0.55,1,0,0.15}
\definecolor[named]{ACMDarkBlue}{cmyk}{1,0.58,0,0.21}

\definecolor[named]{myGreen}{RGB}{31,182,83}
\definecolor[named]{myPurple}{RGB}{117, 112, 179}
\definecolor[named]{myDarkGreen}{RGB}{27,158,119}
\definecolor[named]{myDarkOrange}{RGB}{217,95,2}

\usepackage[framemethod=tikz]{mdframed}
\mdfdefinestyle{mystyle}{%
  rightline=true,
  innerleftmargin=5,
  innerrightmargin=5,
  outerlinewidth=0.8pt,
  topline=true,
  leftline=true,
  bottomline=true,
  skipabove=\topsep
}

\newcommand{\Rone}[1]{#1}
\newcommand{\Rtwo}[1]{#1}
\newcommand{\Rthree}[1]{#1}

\newenvironment{RoneB}{%
}{%
}
\newenvironment{RtwoB}{%
}{%
}

\def\BibTeX{{\rm B\kern-.05em{\sc i\kern-.025em b}\kern-.08em
    T\kern-.1667em\lower.7ex\hbox{E}\kern-.125emX}}

\AtBeginDocument{}%
\AtBeginDocument{}%
\AtBeginDocument{}%

\pgfplotsset{compat=1.18} 
\begin{document}


\title{Optimizing Context-Enhanced Relational Joins}

\author{\IEEEauthorblockN{Viktor Sanca}
\IEEEauthorblockA{
\textit{EPFL}\\
viktor.sanca@epfl.ch}
\and
\IEEEauthorblockN{Manos Chatzakis\IEEEauthorrefmark{2}}
\IEEEauthorblockA{
\textit{EPFL}\\
emmanouil.chatzakis@epfl.ch}
\and
\IEEEauthorblockN{Anastasia Ailamaki}
\IEEEauthorblockA{
\textit{EPFL}\\
anastasia.ailamaki@epfl.ch}
}

\maketitle
\begingroup\renewcommand\thefootnote{\IEEEauthorrefmark{2}}
\footnotetext{Author contributed during an internship at DIAS lab, EPFL.}
\endgroup


\begin{abstract}
\input{00-abstract/main.tex}
\end{abstract}

\input{Commands/selectivities_graph}    
\input{Commands/sequence_graph}         
\input{Commands/cumulative_graph}       
\input{Commands/sel_graph_qcs}          
\input{Commands/sel_graph_qvs}          
\input{Commands/sel_qvs_qcs_graph}      


\section{Introduction} 
\import{01-introduction/}{main.tex}

\section{Motivation and Preliminaries} 
\import{02-motivation/}{main.tex}

\section{Context-Enhanced Relational Join Operator} 
\import{03-model-operator-interaction/}{main.tex}

\section{Logical Optimization} 
\import{04-logical-optimization/}{main.tex}

\section{Physical Optimization} 
\import{05-physical-optimization/}{main.tex}

\section{Evaluation} 
\import{06-evaluation/}{main.tex}

\section{Related Work} 
\import{07-related-work/}{main.tex}

\section{Conclusion} 
\import{08-conclusion/}{main.tex}


\bibliographystyle{IEEEtran}
\bibliography{references.bib}

\end{document}

%% file: 00-abstract/main.tex
Collecting data, extracting value, and combining insights from relational and context-rich sources of many modalities in data processing pipelines presents a challenge for traditional relational DBMS. While relational operators allow declarative and optimizable query specification, they are limited to unsuitable data transformations for capturing or analyzing context.
On the other hand, representation learning models can map context-rich data into embeddings, allowing machine-automated context processing but requiring imperative data transformation integration with the analytical query.\\
To bridge this dichotomy, we present a context-enhanced relational join and introduce an embedding operator composable with relational operators. This enables hybrid relational and context-rich vector data processing, with algebraic equivalences compatible with relational algebra and corresponding logical and physical optimizations. We investigate model-operator interaction with vector data processing and study the characteristics of the $\Epsilon$-join operator. We demonstrate the hybrid context-enhanced relational join operators with vector embeddings \Rone{and evaluate it against a vector database approach}. \Rthree{We show step-by-step the impact of logical and physical optimizations, which result in orders of magnitude execution time improvement resulting in tensor join formulation and outline the performance tradeoffs and case of using scan-based processing against vector indexes.}

%% file: Commands/selectivities_graph.tex
\newcommand{\plotSel}[4]{
\usetikzlibrary{calc}

\centering
\begin{tikzpicture}
\begin{axis}[
    name=sequence,
    width=#2,
    height=#3,
    legend style={at={(0.5,1)},draw=none,
        anchor=south,
        legend columns=#4,
        inner sep=0pt, outer sep=0pt
    },
    axis lines = left,
    axis x line*=bottom,
    axis y line*=left,
    xlabel={Query Sequence},
    xlabel style={align=center},
    ylabel style={align=center},
    ylabel=Selectivity \%,
    ylabel absolute, every axis y label/.append style={yshift=-1.25em},
    line width=1.5pt
]
\pgfplotstableread[col sep=comma]{#1}\vvA
\pgfplotstabletranspose[header=true, colnames from=sequence, input colnames to=seq]\vA\vvA

\addplot+[dash dot,mark=none,ACMDarkBlue] table [x=seq, y=selectivity] {\vA};\addlegendentry{Regular Query}
\addplot+[ultra thin,mark=none,ACMBlue] table [x=seq, y=selectivity] {\vA};\addlegendentry{Online Sampling}
\addplot+[solid,mark=none,myDarkGreen] table [x=seq, y=selectivity laqy] {\vA};\addlegendentry{LAQy $\Delta$}

\end{axis}
\end{tikzpicture}

}

\newcommand{\plotSelNL}[4]{
\usetikzlibrary{calc}

\centering
\begin{tikzpicture}
\begin{axis}[
    name=sequence,
    width=#2,
    height=#3,
    legend style={at={(0.5,1)},draw=none,
        anchor=south,
        legend columns=#4,
        inner sep=0pt, outer sep=0pt
    },
    axis lines = left,
    axis x line*=bottom,
    axis y line*=left,
    xlabel={Query Sequence},
    xlabel style={align=center},
    ylabel style={align=center},
    ylabel=Selectivity \%,
    ylabel absolute, every axis y label/.append style={yshift=-1.25em},
    line width=1.5pt
]
\pgfplotstableread[col sep=comma]{#1}\vvA
\pgfplotstabletranspose[header=true, colnames from=sequence, input colnames to=seq]\vA\vvA

\addplot+[dash dot,mark=none,ACMDarkBlue] table [x=seq, y=selectivity] {\vA};
\addplot+[ultra thin,mark=none,ACMBlue] table [x=seq, y=selectivity] {\vA};
\addplot+[solid,mark=none,myDarkGreen] table [x=seq, y=selectivity laqy] {\vA};

\end{axis}
\end{tikzpicture}

}

%% file: Commands/sequence_graph.tex
\newcommand{\plotSeq}[3]{
\usetikzlibrary{calc}

\centering
\begin{tikzpicture}
\begin{axis}[
    name=sequence,
    width=#2,
    height=#3,
    legend style={at={(0.5,1)},draw=none,
        anchor=south,
        legend columns=-1,
        inner sep=0pt, outer sep=0pt
    },
    y filter/.code=\pgfmathparse{##1/1000}, 
    axis lines = left,
    axis x line*=bottom,
    axis y line*=left,
    xlabel={Query Sequence},
    xlabel style={align=center},
    ylabel style={align=center},
    ylabel=Execution Time (s),
    line width=1.5pt
]

\pgfplotstableread[col sep=comma]{#1}\vvA
\pgfplotstabletranspose[header=true, colnames from=sequence, input colnames to=seq]\vA\vvA

\addplot+[very thick,solid,mark=none,myDarkGreen] table [x=seq, y=laqy] {\vA};\addlegendentry{LAQy}
\addplot+[ultra thick,dotted,mark=none,ACMRed] table [x=seq, y=scan] {\vA};\addlegendentry{Scan}
\addplot+[ultra thin,mark=none,ACMDarkBlue] table [x=seq, y=online] {\vA};\addlegendentry{Online}
\addplot+[loosely dash dot,mark=none,ACMPurple] table [x=seq, y=exact] {\vA};\addlegendentry{Exact}

\end{axis}
\end{tikzpicture}

}

\newcommand{\plotSeqNL}[3]{
\usetikzlibrary{calc}

\centering
\begin{tikzpicture}
\begin{axis}[
    name=sequence,
    width=#2,
    height=#3,
    legend style={at={(0.5,1)},draw=none,
        anchor=south,
        legend columns=-1,
        inner sep=0pt, outer sep=0pt
    },
    y filter/.code=\pgfmathparse{##1/1000}, 
    axis lines = left,
    axis x line*=bottom,
    axis y line*=left,
    xlabel={Query Sequence},
    xlabel style={align=center},
    ylabel style={align=center},
    ylabel=Execution Time (s),
    line width=1.5pt
]

\pgfplotstableread[col sep=comma]{#1}\vvA
\pgfplotstabletranspose[header=true, colnames from=sequence, input colnames to=seq]\vA\vvA

\addplot+[very thick,solid,mark=none,myDarkGreen] table [x=seq, y=laqy] {\vA};
\addplot+[ultra thick,dotted,mark=none,ACMRed] table [x=seq, y=scan] {\vA};
\addplot+[ultra thin,mark=none,ACMDarkBlue] table [x=seq, y=online] {\vA};
\addplot+[loosely dash dot,mark=none,ACMPurple] table [x=seq, y=exact] {\vA};

\end{axis}
\end{tikzpicture}

}

%% file: Commands/cumulative_graph.tex
\newcommand{\plotCum}[3]{
\usetikzlibrary{calc}

\centering
\begin{tikzpicture}
\begin{axis}[
    name=sequence,
    width=#2,
    height=#3,
    legend style={at={(0.5,1)},draw=none,
        anchor=south,
        legend columns=-1,
        inner sep=0pt, outer sep=0pt
    },
    y filter/.code=\pgfmathparse{##1/1000}, 
    axis lines = left,
    axis x line*=bottom,
    axis y line*=left,
    xlabel={Query Sequence},
    xlabel style={align=center},
    ylabel style={align=center},
    ylabel=Execution Time (s),
    line width=1.5pt
]
\pgfplotstableread[col sep=comma]{#1}\vvA
\pgfplotstabletranspose[header=true, colnames from=sequence, input colnames to=seq]\vA\vvA

\addplot+[very thick,solid,mark=none,myDarkGreen] table [x=seq, y=cumul. laqy] {\vA};\addlegendentry{LAQy}
\addplot+[ultra thick,dotted,mark=none,ACMRed] table [x=seq, y=cumul. scan] {\vA};\addlegendentry{Scan}
\addplot+[ultra thin,mark=none,ACMDarkBlue] table [x=seq, y=cumul. online] {\vA};\addlegendentry{Online}
\addplot+[loosely dash dot,mark=none,ACMPurple] table [x=seq, y=cumul. exact] {\vA};\addlegendentry{Exact}

\end{axis}
\end{tikzpicture}

}

\newcommand{\plotCumNL}[3]{
\usetikzlibrary{calc}

\centering
\begin{tikzpicture}
\begin{axis}[
    name=sequence,
    width=#2,
    height=#3,
    legend style={at={(0.5,1)},draw=none,
        anchor=south,
        legend columns=-1,
        inner sep=0pt, outer sep=0pt
    },
    y filter/.code=\pgfmathparse{##1/1000}, 
    axis lines = left,
    axis x line*=bottom,
    axis y line*=left,
    xlabel={Query Sequence},
    xlabel style={align=center},
    ylabel style={align=center},
    ylabel=Execution Time (s),
    line width=1.5pt
]
\pgfplotstableread[col sep=comma]{#1}\vvA
\pgfplotstabletranspose[header=true, colnames from=sequence, input colnames to=seq]\vA\vvA

\addplot+[very thick,solid,mark=none,myDarkGreen] table [x=seq, y=cumul. laqy] {\vA};
\addplot+[ultra thick,dotted,mark=none,ACMRed] table [x=seq, y=cumul. scan] {\vA};
\addplot+[ultra thin,mark=none,ACMDarkBlue] table [x=seq, y=cumul. online] {\vA};
\addplot+[loosely dash dot,mark=none,ACMPurple] table [x=seq, y=cumul. exact] {\vA};

\end{axis}
\end{tikzpicture}

}

%% file: Commands/sel_graph_qcs.tex
\newcommand{\plotSelQCSMicro}[1]{
\usetikzlibrary{calc}

\centering
\begin{tikzpicture}
\begin{axis}[
    name=sequence,
    width=0.9\linewidth,
    height=.45\columnwidth,
    legend style={at={(0.5,1)},draw=none,
        anchor=south,
        legend columns=2,
        inner sep=0pt, outer sep=0pt
    },
    y filter/.code=\pgfmathparse{##1/1000}, 
    axis lines = left,
    axis x line*=bottom,
    axis y line*=left,
    xlabel={Selectivity \%},
    xlabel style={align=center},
    ylabel style={align=center},
    ylabel={Execution Time [s]},
    ylabel absolute, every axis y label/.append style={yshift=-1.25em},
    line width=1.5pt
]
\pgfplotstableread[col sep=comma]{#1}\vvA
\pgfplotstabletranspose[header=true, colnames from=sel, input colnames to=sel]\vA\vvA


\addplot+[ultra thin,mark=none,color=ACMDarkBlue] table [x=sel, y=qcs-50] {\vA};\addlegendentry{|50|-stratified}
\addplot+[dashdotted,mark=none,color=ACMDarkBlue] table [x=sel, y=gb-qcs-50] {\vA};\addlegendentry{|50|-groupby}

\addplot+[dotted,mark=none,color=ACMBlue] table [x=sel, y=qcs-4950] {\vA};\addlegendentry{|4950|-stratified}
\addplot+[solid,mark=none,color=ACMBlue] table [x=sel, y=gb-qcs-4950] {\vA};\addlegendentry{|4950|-groupby}

\end{axis}
\end{tikzpicture}

}

%% file: Commands/sel_graph_qvs.tex
\newcommand{\plotSelQVSMicro}[1]{
\usetikzlibrary{calc}

\centering
\begin{tikzpicture}
\begin{axis}[
    name=sequence,
    width=0.9\linewidth,
    height=.45\columnwidth,
    legend style={at={(0.5,1)},draw=none,
        anchor=south,
        legend columns=2,
        inner sep=0pt, outer sep=0pt
    },
    y filter/.code=\pgfmathparse{##1/1000}, 
    axis lines = left,
    axis x line*=bottom,
    axis y line*=left,
    xlabel={Selectivity \%},
    xlabel style={align=center},
    ylabel style={align=center},
    ylabel={Execution Time [s]},
    ylabel absolute, every axis y label/.append style={yshift=-1.25em},
    line width=1.5pt
]
\pgfplotstableread[col sep=comma]{#1}\vvA
\pgfplotstabletranspose[header=true, colnames from=sel, input colnames to=sel]\vA\vvA



\addplot+[ultra thin,mark=none,color=myGreen] table [x=sel, y=qvs-50] {\vA};\addlegendentry{|50|-stratified}
\addplot+[dashdotted,mark=none,color=myGreen] table [x=sel, y=gb-qvs-50] {\vA};\addlegendentry{|50|-groupby}

\addplot+[dotted,mark=none,color=myDarkGreen] table [x=sel, y=qvs-4950] {\vA};\addlegendentry{|4950|-stratified}
\addplot+[solid,mark=none,color=myDarkGreen] table [x=sel, y=gb-qvs-4950] {\vA};\addlegendentry{|4950|-groupby}

\end{axis}
\end{tikzpicture}

}

\newcommand{\plotSelQVSMicroNL}[1]{
\usetikzlibrary{calc}

\centering
\begin{tikzpicture}
\begin{axis}[
    name=sequence,
    width=0.9\linewidth,
    height=.45\columnwidth,
    legend style={at={(0.5,1)},draw=none,
        anchor=south,
        legend columns=2,
        inner sep=0pt, outer sep=0pt
    },
    y filter/.code=\pgfmathparse{##1/1000}, 
    axis lines = left,
    axis x line*=bottom,
    axis y line*=left,
    xlabel={Selectivity \%},
    xlabel style={align=center},
    ylabel style={align=center},
    ylabel={Execution Time [s]},
    ylabel absolute, every axis y label/.append style={yshift=-1.25em},
    line width=1.5pt
]
\pgfplotstableread[col sep=comma]{#1}\vvA
\pgfplotstabletranspose[header=true, colnames from=sel, input colnames to=sel]\vA\vvA



\addplot+[ultra thin,mark=none,color=ACMDarkBlue] table [x=sel, y=qvs-50] {\vA};
\addplot+[dashdotted,mark=none,color=ACMDarkBlue] table [x=sel, y=gb-qvs-50] {\vA};

\addplot+[dotted,mark=none,color=ACMBlue] table [x=sel, y=qvs-4950] {\vA};
\addplot+[solid,mark=none,color=ACMBlue] table [x=sel, y=gb-qvs-4950] {\vA};

\end{axis}
\end{tikzpicture}

}

%% file: Commands/sel_qvs_qcs_graph.tex
\newcommand{\plotSelQCSQVSMicro}[1]{
\usetikzlibrary{calc}

\centering
\begin{tikzpicture}
\begin{axis}[
    name=sequence,
    width=0.9\linewidth,
    height=.65\columnwidth,
    legend style={at={(0.5,1)},draw=none,
        anchor=south,
        legend columns=2,
        inner sep=0pt, outer sep=0pt
    },
    y filter/.code=\pgfmathparse{##1/1000}, 
    axis lines = left,
    axis x line*=bottom,
    axis y line*=left,
    xlabel={Selectivity \%},
    xlabel style={align=center},
    ylabel style={align=center},
    ylabel={Execution Time [s]},
    ylabel absolute, every axis y label/.append style={yshift=-1.25em},
    line width=1.5pt
]
\pgfplotstableread[col sep=comma]{#1}\vvA
\pgfplotstabletranspose[header=true, colnames from=sel, input colnames to=sel]\vA\vvA



\addplot+[ultra thin,mark=none,color=ACMDarkBlue] table [x=sel, y=qvs-50] {\vA};\addlegendentry{|50|-stratified}
\addplot+[dashdotted,mark=none,color=ACMDarkBlue] table [x=sel, y=gb-qvs-50] {\vA};\addlegendentry{|50|-groupby}

\addplot+[dotted,mark=none,color=ACMBlue] table [x=sel, y=qvs-4950] {\vA};\addlegendentry{|4950|-stratified}
\addplot+[solid,mark=none,color=ACMBlue] table [x=sel, y=gb-qvs-4950] {\vA};\addlegendentry{|4950|-groupby}

\end{axis}
\end{tikzpicture}

}

%% file: 01-introduction/main.tex
\label{sec:introduction}
Relational databases allow declarative query specification and abstractions for logical and physical query plan optimizations. These optimizations include operator reordering via algebraic equivalences and heuristics and instantiating operators for resource-efficient and hardware-conscious execution on modern hardware. This allows ad-hoc query specification by abstracting out significant implementation details from the user and end-to-end optimization. As the main goal of relational analytical databases is to provide abstractions to large-scale processing and extraction of value from the data of interest, relational databases are designed for data types where a procedural way to process the data is possible to precisely specify, such as aggregating numerical values or processing strings with a well-specified pattern. 

Still, many data sources are unsuitable for processing in a relational database and are typically only stored serialized in binary formats. These include documents, text, images, and other data sources of increasing value, driven by the advent of the Internet and mobile devices and services such as social media. Such data has a lot of human-understandable contexts: the contents and number of objects in an image, the semantics of a string despite alternative spellings, typos, or tenses, all of which make this task impractical if not impossible to specify in traditional relational data analytics. 

On the other hand, advancements in artificial intelligence and machine learning have allowed increasingly complex machine reasoning and performance in analyzing context-rich data such as images or text. Models such as BERT~\cite{DBLP:conf/naacl/DevlinCLT19} and GPT~\cite{DBLP:conf/nips/BrownMRSKDNSSAA20} allow natural language processing, ResNet~\cite{DBLP:conf/cvpr/HeZRS16} object localization and detection, often available as Foundation Models~\cite{DBLP:journals/corr/abs-2108-07258} that are trained on web-scale data and further customizable and re-trainable for the particular task. To use those models, often based on Transformer architecture~\cite{DBLP:conf/nips/VaswaniSPUJGKP17}, analysts would instantiate the particular model, input data, and collect the output, using frameworks such as Tensorflow~\cite{abadi2016tensorflow} or Pytorch~\cite{paszke2019pytorch}, often in an isolated, task-specific setting. 
With the proliferation of embedding-based analytics, vector databases have recently gained traction, offering vector processing \Rthree{based on index structures~\cite{DBLP:journals/tbd/JohnsonDJ21}} with limited integration with traditional relational analytics or available operations over data.

\begin{figure}
  \centering
  \includegraphics[width=0.8\linewidth]{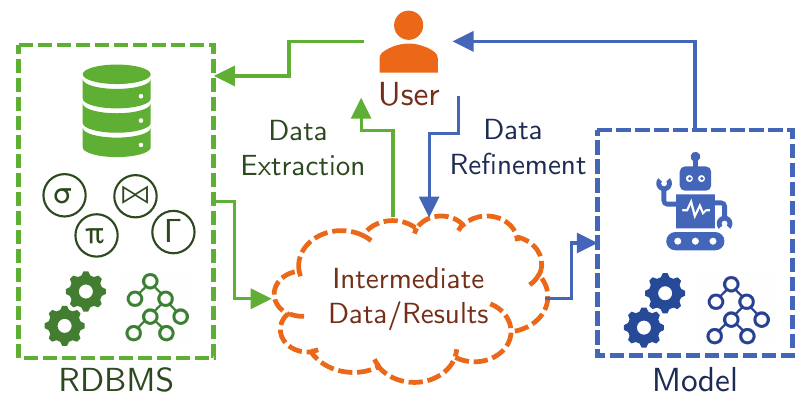}
  \caption{Problem: Model-RDBMS data analysis requires user expertise, imperative tasks, and data movement specification.}
  \label{fig:motivation}
  \vspace{-15pt}
\end{figure}


While machine learning models can transform context-rich, multi-modal data into embeddings, coordinating the models and data processing pipelines is manual and imperative. Suppose an analyst wanted to analyze and extract insights from an RDBMS and use some data as input to the models as in \autoref{fig:motivation}, which may be again input to an analytical query and models in a more complex analytical processing pipeline. Combining the data from social media feeds with user reviews, transactions, and analytics in an online retailer case results in complex, hybrid data processing pipelines. The data of interest and value is both relational and vector-based, not fitting fully to cases outlined in TPC-H, TPC-DS, or SSB~\cite{o2007star} benchmarks, \Rthree{or vector search-aimed ANN-Benchmarks~\cite{DBLP:journals/is/AumullerBF20}}.

\begin{figure}
  \centering
  \includegraphics[width=\linewidth]{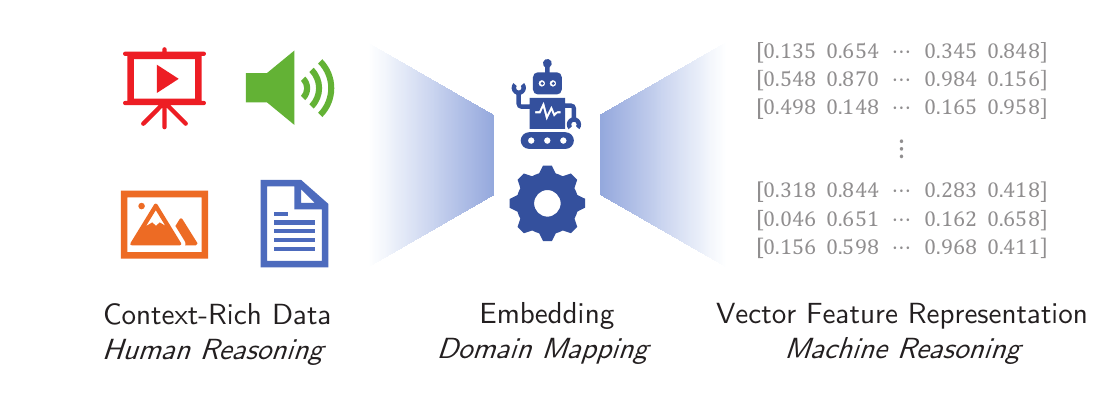}
  \caption{Enabler: models embed context-rich data into common tensor representation, enabling automated processing.}
  \label{fig:enabler}
  \vspace{-10pt}
\end{figure}

With two independent components, RDBMS with relational data and Model with vector data, the user must be an expert to fine-tune the queries, potentially perform data integration, correctly deploy and scale the queries to the hardware, orchestrate data movement, and specify the correct operator orders to prevent negatively impacting the performance. 
\Rthree{To keep the principle of \textit{data independence}~\cite{DBLP:journals/cacm/Codd70}, hiding the implementation details from the user, we propose building on top of } decades of research and engineering in query optimization and execution engines allow hiding this complexity and is the key motivation to extend and make relational algebra the basis of emerging methods in multi-modal and context-rich processing.
%
Tightly integrated, expressive, and optimizable, hybrid vector-relational data management is part of our broader effort for the next generation of context-rich analytical engines~\cite{DBLP:conf/icde/SancaA23}, enabled by embedding models that transform the context-rich data into tensors as a common intermediate data representation (\autoref{fig:enabler}). A separation of concerns is established: the model handles modality, data context, and semantics; the analytical engine optimizes and processes context-free data and tensors via defined operators.

\Rthree{As the data of interest can be a mix of relational and vector data, this results in} operators having different physical and logical properties, \Rthree{which, naively built on top of relational operators, yield in suboptimal performance}. 
In this work, we investigate the case of context-enhanced join operator, which takes place over vector embeddings and relational data, and:
\begin{itemize}
    \item Motivate and propose the capabilities of a context-enhanced join operator in \autoref{sec:motivation}, and introduce and formalize the relational operator extension in \autoref{sec:model-operator-interaction},
    \item Analyze the suitability of traditional join operator for the task of vector data processing, and propose a cost model, logical optimizations, and an alternative efficient tensor formulation for parallel execution of a join operator for processing neural embeddings in \autoref{sec:logical-optimization},
    \item Evaluate the physical and hardware optimizations we propose in \autoref{sec:physical-optimization}, and benchmark the operator implementation and characteristics in \autoref{sec:evaluation}, showing the over orders of magnitude impact on the execution time and the importance of both logical and physical optimizations of vector-based join operations, \Rone{and comparing the tradeoffs for using a state-of-the-art index in a vector database}.
\end{itemize}
%
%
%
%


%% file: 02-motivation/main.tex
\label{sec:motivation}

\begin{figure}
  \centering
  \includegraphics[width=\linewidth]{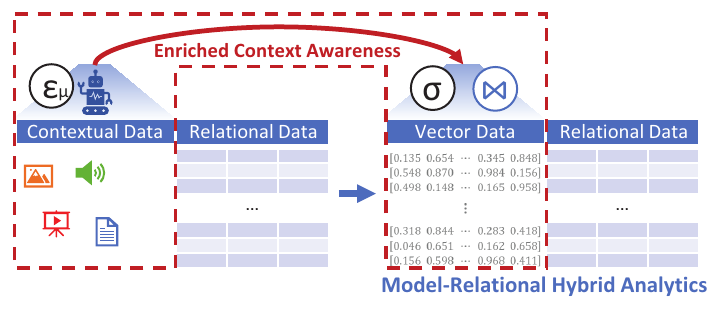}
  \caption{Context-enhanced, model-relational analytics.} 
  \label{fig:design_goal}
  \vspace{-15pt}
\end{figure}

There have been significant efforts to enable machine-automated understanding of context-rich data.
\Rone{We first define key concepts, starting from \textit{context-rich data}, which contains information and metadata, often human-understandable but not directly machine-understandable, such as word synonyms or understanding that two images are identical. The key idea behind neural embeddings is that the machine learning \textit{model} ($\mu$) learns how to transform the input data domain into high-dimensional vector space (\autoref{fig:enabler}), where relationships between the data can be expressed using linear algebra expressions over vectors. Embedding models ($\Epsilon_\mu$) take a human-understandable context-rich domain and map it into a machine-operable high dimensional vector space representation (\textit{tensors}), where \textit{embeddings} are tensor representations of model inputs.}

\begin{RtwoB}
Embedding models enable data analysis over many data modalities~\cite{DBLP:conf/naacl/DevlinCLT19, DBLP:journals/taslp/KongCIWWP20, DBLP:conf/cvpr/HeZRS16,DBLP:journals/jmlr/RaffelSRLNMZLL20}, which was previously done manually, and was difficult or impossible for applications such as e-commerce~\cite{10.1145/3219819.3219869}, search~\cite{10.1145/3394486.3403305}, Retrieval-Augmented Generation (RAG)~\cite{DBLP:conf/nips/LewisPPPKGKLYR020}, and recommender systems~\cite{zhang2016collaborative}. Vector indexes~\cite{DBLP:journals/tbd/JohnsonDJ21} with vector databases~\cite{DBLP:conf/sigmod/WangYGJXLWGLXYY21} provide a framework for search and retrieval, with inherently similarity-based operations over vector embeddings. This contrasts with traditional relational analytics, where expressions have exact semantics. Unifying both and investigating the tradeoffs in the hybrid space in between is the topic of our study. 
\end{RtwoB}


\subsection{Extended Functionality: Joins Over Contextual Data}

Model-driven embeddings transform data previously opaque to the relational data management system into context-free vectors (\autoref{fig:design_goal}). 
Separating concerns between the model-driven context and uniform vector data representation enables defining expressions over vectors and tensors \Rthree{, maintaining the relational model principle of \textit{data independence}~\cite{DBLP:journals/cacm/Codd70}}.
Our approach outlines a way to join context-rich data such as strings, documents, or images by defining a corresponding model and vector expression as first-class operations alongside relational operators, enabling novel logical and physical optimizations.

\Rtwo{Similarity joins are an important class of operations in data cleaning and integration designed to tolerate the errors and inconsistencies in the data and extend the traditional exact joins. They have a long tradition and many real-world applications, such as entity resolution, duplicate detection, spell-checking, and clustering~\cite{DBLP:journals/fcsc/YuLDF16,DBLP:journals/pvldb/JiangLFL14,DBLP:journals/pvldb/ChenWNC19,DBLP:journals/vldb/SilvaALPA13}, finding place in production platforms such as Microsoft SQL Server Integration
Services~\cite{Microsoft2023FuzzyLookup}, Informatica~\cite{InformaticaDataQuality2024}, Knime~\cite{KNIME2024}, and Talend~\cite{Talend2024}. The proliferation of vector embeddings for contextual data further extends the need for an efficient similarity join, expanding the aforementioned use cases, where we next discuss some novel applications.}

\subsubsection{Semantic-Based Similarity Operations}\label{ssec:semantic}

Instead of having a human-in-the-loop or an expert system that performs dictionary-based or hard-coded rule-based similarity operations\Rtwo{~\cite{DBLP:journals/fcsc/YuLDF16,DBLP:journals/pvldb/ChenWNC19}}, models automate similarity operations over \Rtwo{many data types~\cite{DBLP:journals/taslp/KongCIWWP20,DBLP:conf/naacl/DevlinCLT19,DBLP:conf/cvpr/HeZRS16}, including multi-modal data~\cite{DBLP:journals/jmlr/RaffelSRLNMZLL20}}. A common tensor representation poses similarity joins simply as similarity expressions, such as cosine distance, between the vector-embedded data. The models fine-tune the functionality and context. After embedding the data and providing operators and expressions over tensors such as cosine distance, model-independent operations can be combined with the rest of the relational query plan.

\subsubsection{Online Data Cleaning and Integration}\label{ssec:data-cleaning}



Strings or other context-rich data can be dirty or have rich semantics.\Rtwo{~\cite{DBLP:journals/fcsc/YuLDF16,10.5555/645927.672200}}
If we consider words or sentences, they may have misspellings, alternative spellings, synonyms, or different tenses that all have the same meaning.
Specifying all the rules to unify context is error-prone and difficult. Meanwhile, embeddings can encompass such similarity using representation learning. Therefore, such operators can process such data on the fly without prior cleaning and only the data of interest, relying on embeddings and specified similarity thresholds for data integration and potentially performing post-verification steps.

\subsubsection{Multi-Modal Data Processing}\label{ssec:multi-modal-data}
The data context is opaque to the execution engine, while the model selection and parameters give context and transform the data. By processing context-free tensors, relational engines provide a common optimization framework for multi-modal data ingested and transformed initially by models. 
\Rtwo{A multi-modal similarity join is useful in near-duplicate detection of unlabeled entries against another database, such as in misinformation detection~\cite{AIMeta2024CovidMisinformation} or document tagging~\cite{DBLP:conf/rep4nlp/ChenSPM17}. More generally, as a search query takes a single query as an input, batching many search queries would be equivalent to a join operation for better use of the available parallelism ~\cite{DBLP:conf/sigmod/KesterAI17}.}\\

\subsection{Integrating Vector Embeddings With Relational Operators}
Data management systems support and simplify data processing with research and systems contributions and features such as transactions and concurrency control~\cite{zhang21, DBLP:conf/icde/KemperN11}, auto-tuning~\cite{pavlo17}, hardware-conscious implementations~\cite{DBLP:journals/pvldb/Neumann11, DBLP:journals/pvldb/ChrysogelosKAA19, DBLP:conf/cidr/NeumannF20, DBLP:journals/pvldb/KerstenLKNPB18, DBLP:conf/icde/ZukowskiWB12}, corresponding data structures~\cite{DBLP:conf/sigmod/IdreosZHKG18}, and query optimization with declarative interfaces to abstract out the system complexity from the end-user.
%
Instead of manual intermediate orchestration and system integration to combine and analyze multi-modal and context-rich data involving different systems, data sources, and efficient operator reimplementation, we investigate how to extend traditional relational joins to support model-driven context with minimal system intrusions and build on top of existing judiciously modified abstractions. In particular, this means that the vectors are simply another data type over which expressions and operations can be defined.
This makes index structures designed to store, maintain, and perform similarity search over tensors~\cite{DBLP:journals/tbd/JohnsonDJ21} compatible as physical access method options. 
Similarly, recent work has formulated traditional relational processing over tensors~\cite{DBLP:journals/corr/abs-2211-02753}, where tensor processing platforms are used as analytical RDBMS to benefit from existing implementation while transforming the relational data and operations into tensor representation. 

\begin{figure}
  \centering
  \includegraphics[width=0.85\linewidth]{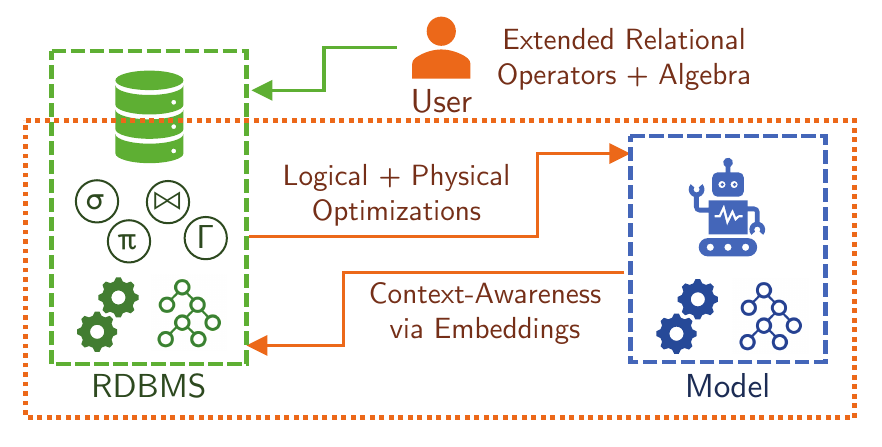}
  \caption{Goal: Hybrid vector-relational operations are declarative transformation primitives amenable to query optimization.} 
  \label{fig:motivation_solution}
  \vspace{-10pt}
\end{figure}

While there has been prior work to integrate model inference and learning with analytical engines~\cite{DBLP:conf/icde/LinWZDLC22, DBLP:journals/pvldb/HellersteinRSWFGNWFLK12}, our goal is complementary as we focus on how we can extend the relational model functionality with contextual data, as illustrated in \autoref{fig:motivation_solution}. Similarly, we expose co-optimization opportunities at logical, physical, and implementation levels and fine-grained system interactions~\cite{DBLP:conf/icde/SancaA23}.
\Rone{As vector databases are typically based on index structures~\cite{DBLP:conf/sigmod/WangYGJXLWGLXYY21}, in this work, we formulate our operator for both index-probes and scans, considering the relational attribute-driven selectivity of analytical queries~\cite{DBLP:conf/sigmod/KesterAI17}.}

\subsection{Holistic Optimization}\label{ssec:holistic-optimization}

Suppose the data of interest are strings and dates stored in an RDBMS. One can consider other context-rich formats stored as binary objects with other relational data. To allow semantic similarity operations, such as matching strings that are synonyms, have misspellings, or different tenses, word embedding models transform strings into vectors, which are then comparable using cosine distance. 
While RDBMS could execute regex-like string expressions, mapping strings to embeddings allows capturing broader classes of similarity within a model.
Note that the model can be trained and adapted for different datasets to adjust the notion of similarity, which the analyst selects.

We are interested in joining two tables over strings, where a condition over dates exists, making the queries selective on both tables. 
In a declarative setting, query specification requires embedding model information and the join condition expression, and the selectivity information from the relational column needs to propagate before the embeddings. Otherwise, the whole interaction may result in the user eagerly materializing all the data as in~\autoref{fig:motivation}, performing expensive embedding, and only then filtering.

Physical optimization must address this interaction and account for the tensor data format and the expressions suitable for comparing high-dimensional data. For example, while an equi-join over tensors could be implemented as a hash-join, more practical embedding comparisons, such as cosine distance, require algorithms such as nested-loop join for pair-wise comparisons and consider the join, operation, and model data structure access patterns in the algorithm and cost model. \Rthree{For example, if the queries are selective and a vector index exists, should we select it or use an exhaustive scan for the operation~\cite{DBLP:conf/sigmod/KesterAI17}.}

Finally, from a hardware-conscious perspective, using many-dimensional vectors with relational operators designed and optimized for single-dimensional numerical data and judicious use of caches and memory hierarchy demands novel tradeoffs. A 100-dimensional tensor embedding will change the caching and execution patterns of traditional algorithms, and model embedding can incur computational or data access costs at a critical path of execution. Designing hardware-conscious algorithms represents a direction driven by novel model-database interactions.
We demonstrate join operator optimizations (\autoref{fig:motivation_solution}), step-by-step, from logical and physical optimization that enable efficient execution. 



\textbf{Takeaway} Neural embedding models transform the context-rich, multi-modal data into a common (per-model) tensor representation space. From the perspective of declarative relational processing, models provide separation between data semantics and context-less tensors. Relational operators perform operations such as cosine distance or vector transformations over tensors, amenable to query optimization. We propose novel optimizations and analyze their performance, aware of the new design space.

%% file: 03-model-operator-interaction/main.tex
\label{sec:model-operator-interaction}

In this section, we start with formalizing the proposal of a relational operator extension to declaratively process context-rich data stored along traditional relational data, such as strings and text, that may be stored along with numerical or date attributes. We call this \textit{hybrid model-relational processing}. This enhancement stems from the fact that the contextual data may need to be transformed and processed differently. However, compatibility with relational algebra and optimizations for processing purely relational data is required. Instead of using separate systems and manually orchestrating the data movement for processing using external programs or opaque UDFs, we propose a set of operations needed to express a join based on embedding the original data that is amenable to traditional query optimization.

\subsection{Neural Embeddings}  \label{ssec:neur-emb}
Neural embeddings and representation learning are rich and active research fields in machine learning. Images can be embedded with models such as ResNet~\cite{DBLP:conf/cvpr/HeZRS16}, audio with PANNss~\cite{DBLP:journals/taslp/KongCIWWP20}, and text with Bert~\cite{DBLP:conf/naacl/DevlinCLT19}, word2Vec~\cite{DBLP:journals/corr/abs-1301-3781}, FastText~\cite{DBLP:journals/tacl/BojanowskiGJM17, DBLP:conf/naacl/EdizelPBFGS19}. Foundation Models~\cite{DBLP:journals/corr/abs-2108-07258} offer an increasingly flexible way to specialize large models to a particular use case. It is important to mention those models can be tuned, as they learn representations from the training dataset through transfer learning~\cite{DBLP:conf/naacl/QiSFPN18} (e.g., starting from one of the foundation models) or re-training. In this work, we focus on and experiment with string embedding models. However, as embeddings are generally high-dimensional vectors, once in the embedding domain, the processing of this data is model- and input-data-type-agnostic, and the same principles and optimizations hold.
Processing embedded data allows automating semantic similarity using cosine similarity (or another distance) between the vectors. 
Using vectors necessitates interaction with linear algebra; therefore, the equations below outline definitions of cosine similarity over vectors and matrices. We will use them heavily in logical (\autoref{sec:logical-optimization}) and physical (\autoref{sec:physical-optimization}) optimization phases.

\begin{equation}
\tag{Cosine Similarity}
\label{eq:cosine-sim}
cos(\theta) = \frac{A \cdot B}{\lVert A \rVert\lVert B \rVert} = \frac{\sum\limits_{i=1}^{n}A_i B_i}{\sqrt{\sum\limits_{i=1}^{n}A_i^{2}}\sqrt{\sum\limits_{i=1}^{n}B_i^{2}}}
\end{equation}
\vspace{-15pt}



\subsection{Model-Operator Interaction}

In our example, we focus on context awareness over strings so that common mistakes or semantically similar words are automatically captured. Rather than imposing user to strictly specify the rules for string similarity or clean the data ahead of time, we enable words such as (\texttt{barbecue}, \texttt{barbecues}, \texttt{bbq}, \texttt{barbicue}, \texttt{grilling}) that have similar semantics, to automatically be used with relational operator predicates without prior user intervention. The user should only specify the embedding model and a threshold distance parameter over cosine similarity calculation (\autoref{eq:cosine-sim}). Instead of comparing two strings in their original domain, they are embedded. If the cosine similarity $cos(\theta)$ is larger than the specified threshold, the two strings are similar and should be matched. This avoids manual string processing and combining techniques, such as Locality Sensitive Hashing, individually limited to capturing only features such as misspellings.

A context-aware operator is supplemented with an embedding model ($\mu$). In this case, when an operator receives strings, it embeds them and then performs the requested processing in the vector domain. Models can be selected based on the analyst's needs, while often having desirable properties such as the capability of training and adapting to the desired similarity context.
This interaction opens up design and optimization choices, such as how to mask or minimize the cost of embedding/model and overlap it with operator execution. 
We capture this interaction through relational algebra (\autoref{ssec:rel-ops}) and a cost model (\autoref{sec:logical-optimization}) to allow holistic integration with the remainder of the query plan.

\subsection{Relational Operators and Algebra} \label{ssec:rel-ops}

We introduce the embedding operator ($\Epsilon$) using a model ($\mu$), and relational algebra equivalences over selection ($\sigma$) and $\theta$-join ($\bowtie_{\theta}$) operations, compatible with traditional relational algebra definitions.

\textit{Selection} operation applies predicate $\theta$ over input tuples and returns only the tuples that satisfy the condition.
\begin{equation}
\tag{Selection with predicate $\theta$}
\sigma_{\theta}(R) = \{ t \in R,\ \theta(t) \}
\end{equation}

To change the domain of input data, we allow mapping the input tuples (or a projection over the tuples for simple notation) using a model ($\mu$) into vector space using embedding ($\Epsilon$) operation. 
\begin{equation}
\tag{Embedding with model $\mu$}
\Epsilon_{\mu}(R) = \{ t \in R,\ t \mapsto \mu(t) \}
\end{equation}

For completeness and decoding of the embeddings and retrieving the context-rich data, an inverse operation $\Epsilon^{-1}$ should also be defined, which is the standard component of encoder-decoder architectures, and semantically correct only for the same model $\mu$. \Rone{If the model does not have a decoder to recover the original data R, a lookup table mechanism can maintain the object-embedding mapping via unique IDs.}
%
\begin{equation}
\label{eq:decode}
\tag{Decoding with model $\mu$}
\Epsilon^{-1}_{\mu}(\Epsilon_{\mu}(R)) = R
\end{equation}

Combining embedding with selection allows the processing of tuples with a mixture of data formats. Some attributes may have relational predicates, and some may require embedding and predicates using different metrics (such as cosine distance). Predicate pushdown and operation reordering can happen as soon as the attributes that predicates operate over are available.
\begin{equation}
\tag{$\Epsilon$-Selection}
\sigma_{\Epsilon, \mu, \theta}(R) 
\Leftrightarrow 
\sigma_{\theta}(\Epsilon_{\mu}(R))
\Leftrightarrow 
\sigma_{\theta_{\Epsilon}}(\Epsilon_{\mu}(\sigma_{\theta_{R}}(R))))
\end{equation}

\textit{Join} operation takes two relations and joins them over specified attributes using specified predicate conditions ($\theta$-join).
\begin{equation}
\tag{Cartesian Product}
R \times S = \{ (r,s),\ r \in R \wedge s \in S \}
\end{equation}
Joins are amenable to predicate pushdowns and reordering.
\begin{equation}
\tag{$\theta$-Join Generalization}
R \bowtie_{\theta} S 
\Leftrightarrow 
\sigma_{\theta}(R \times S)  
\end{equation}

We introduce embeddings to the generalized join definition and provide equivalences. Embeddings can be observed as a special projection operation that changes the domain.
\begin{equation}
\tag{$\Epsilon$-$\theta$-Join}
\begin{split}
R \bowtie_{\Epsilon, \mu, \theta} S 
\Leftrightarrow 
\sigma_{\Epsilon, \mu, \theta}(R \times S)\\
\Leftrightarrow\\
\sigma_{\theta}(\Epsilon_{\mu}(R) \times \Epsilon_{\mu}(S))
\Leftrightarrow
\Epsilon_{\mu}(R) \bowtie_{\theta} \Epsilon_{\mu}(S) 
\end{split}
\end{equation}


\begin{figure}
  \centering
  \includegraphics[width=0.85\linewidth]{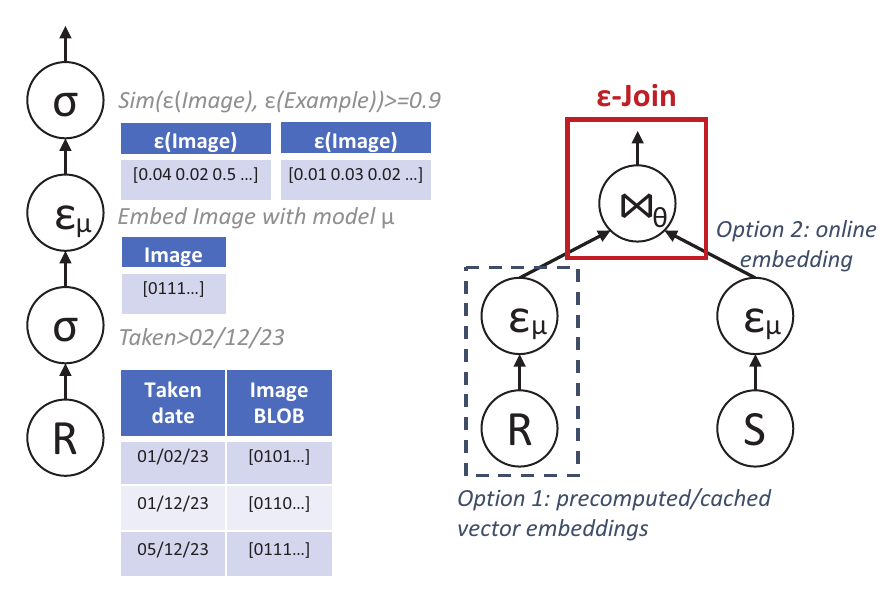}
  \caption{Hybrid vector-relational query example, and the join operator which is the focus of the optimizations in this paper.}
  \label{fig:operators}
  \vspace{-15pt}
\end{figure}

\textbf{Takeaway}
We formulate the context-enhanced operators by extending relational operators and algebra to allow declarative integration of embedding models with relational engines and optimizers. A hybrid setting enables declarative and systematic logical and physical optimizations, as depicted in the simple query in \Cref{fig:operators}, providing semantic awareness using embeddings to separate concerns between models and engines.

%% file: 04-logical-optimization/main.tex
\label{sec:logical-optimization}

\begin{figure*}[ht]
  \centering
  \includegraphics[width=0.75\linewidth]{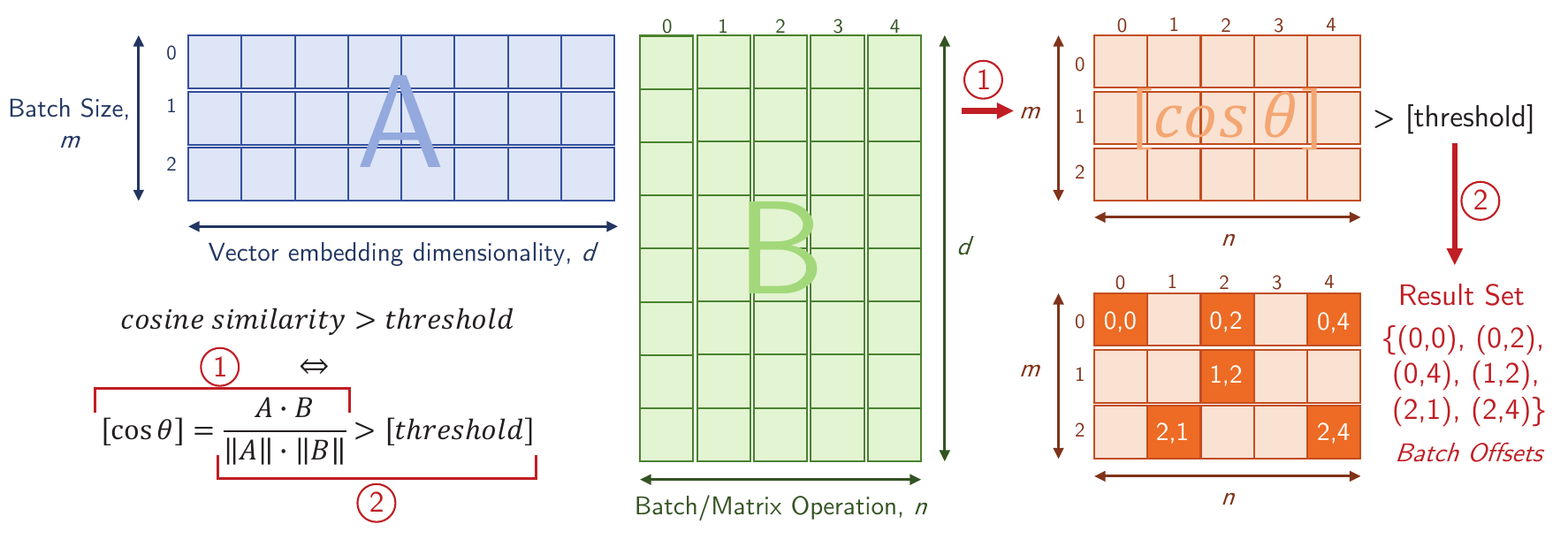}
  \caption{Matrix formulation of $\Epsilon$-join allows scalable and cache-efficient execution over high-dimensional embeddings.}
  \label{fig:tensor}
  \vspace{-15pt}
\end{figure*}

Starting from the extended algebra and operators, we present the logical optimization driven by model-operator interaction and tensors as a common intermediate data representation. In contrast to traditional optimizations and relational operator cost models, the two factors are different. First, since models may be on the critical path of execution, model embedding data access or computation time must be considered in addition to the relational operator's data access and processing cost. 
Second, \Rone{embeddings extend the notion of \textit{atomic} data types prescribed by \textit{1$^{st}$ normal form}~\cite{DBLP:persons/Codd71a}, satisfying the condition that an \textit{"atomic data cannot be decomposed into smaller pieces by the DBMS (excluding certain special functions)"}~\cite{10.5555/77708}. Embeddings are not structured data but should be observed and processed atomically by the DBMS.}


\subsection{Cost Model}


We outline the abstract cost model for the context-enhanced similarity selection and join operations. For joins, we investigate the first available strategy: nested-loop join (NLJ)\Rone{~\cite{DBLP:journals/fcsc/YuLDF16}}. 
We note that we focus on evaluating exact algorithms in our study. Since the distance we use is cosine-similar, hash-based approaches would yield approximate solutions similar to locality-sensitive hashing. 
If we were to use equi-joins, it would be possible to use traditional hash-joins, but there would be no benefit from using embeddings.
Still, nested loop joins can be formulated with good cache-locality, an important performance factor (\autoref{sec:evaluation}) that does not incur random access over high dimensional data as every vector needs to be pair-wise compared using cosine distance.

We outline the abstract cost model for selection and join below, where $R$ and $S$ are relations, $|R|$ is the cardinality of relation $R$, $A$ represents the data access cost, $M$ represents the model cost, and $C$ is the computation cost \Rone{of the operation}. \Rone{As the model, computation, and access cost vary relative to the particular architecture and DBMS, the cost model should be parametrized based on their mutually normalized relative performance~\cite{DBLP:conf/sigmod/KesterAI17,DBLP:journals/pvldb/HilprechtB22}}.

Selection is an operation where input data is scanned, embedded, and the condition is applied over every input tuple, where each tuple incurs access, model, and computation cost (\Rone{when the tuples are only retrieved $C = 0$}): 
\begin{equation}
    \tag{$\Epsilon$-Selection Cost}
    Cost(\sigma_{\Epsilon, \mu, \theta}(R)) = |R| \cdot (A + M + C) 
\end{equation}

By naively extending the Nested-Loop Join (NLJ) operation, it would scan both relations and perform pair-wise condition comparisons. In this implementation, model access would be performed per-processed tuple without considering model-relational interaction, which incurs quadratic model access cost. Considering that embedding models are computationally expensive, the following cost model shows the suboptimality:
\begin{equation}
    \label{nlj-naive}
    \tag{$\Epsilon$-NL Join Cost}
    Cost(R \bowtie_{\Epsilon, \mu, \theta} S) = |R| \cdot |S| \cdot (A + M + C)
\end{equation}

Instead, by considering the characteristics of the nested-loop join, we observe that tuple embedding needs to happen only once per tuple from both relations. This can be performed as a precursor to the join operation or as a lazy embedding and data materialization strategy. By observing this behavior, the join results in a linear model cost with prefetching:
\begin{equation}
    \label{nlj-prefetch}
    \tag{$\Epsilon$-NLJ Prefetch Optimization}
    Cost(R \bowtie_{\Epsilon, \mu, \theta} S) = |R| \cdot |S| \cdot (A + C) + (|R| + |S|) \cdot M 
\end{equation}

This optimization is significant, as the model cost can span from random access to a lookup table (several times slower than sequential scan) to expensive computations over deep neural networks (data transfer and computation). 
From another perspective, if machine learning models are used as-a-service and paid for per embedding~\cite{DBLP:conf/vldb/SancaA23a}, this cost model conversely results in monetary savings compared to the initial implementation.
Expressing embeddings as relational operator extensions allows logical optimization to occur in conjunction with other operators in the hybrid relational-embedding pipeline (selection pushdown, join reordering), such that the cardinality of the most costly part of the query plan will be reduced without explicit user intervention or knowledge about specific interactions given a relational operator.


\begin{RoneB}
\subsection{Index Join Formulation}
Current vector databases rely on index structures~\cite{DBLP:journals/tbd/JohnsonDJ21,10.1109/TPAMI.2018.2889473} that accelerate vector search. They avoid exhaustive cross-product per-vector evaluation at the cost of approximate results and random access patterns. If an index exists on S,  the index probe cost is denoted as $I_probe$, the join cost becomes:

\begin{equation}
    \label{index-join-prefetch}
    \tag{$\Epsilon$-Index Join Cost}
    Cost(R \bowtie_{\Epsilon, \mu, \theta} S) = |R| \cdot I_{probe}(S) \cdot (A + C) 
\end{equation}

{\Large
\renewcommand{\arraystretch}{1.5}
\noindent
\begin{table}
    \centering
    \begin{RoneB}
    \begin{tabular}{c|c|c}
        & \textbf{Scan Join} & \textbf{Index Join}\\
        \hline
        \textbf{Accuracy} & Exact & Approximate \\
        \hline
        \textbf{Filtering} & Full Relational & Vector Similarity \& Pre-Filtering\\
        \hline
        \textbf{Cost} & Compute \& Scan & Build \& Compute \& Probe \\
        \hline
        \textbf{Flexibility} & Any Expression & Limited, Construction-Time Distance
        \vspace{-5pt}
    \end{tabular}
    \caption{\Rone{Index versus scan-based vector join operator.}}
    \vspace{-15pt}
    \label{tab:comparison}
    \end{RoneB}
\end{table}
\vspace{-15pt}
}
 
Probing the index removes the need for a full cross-product by pruning out the search space at construction time, reducing the necessary computation.
This assumes an index on the particular similarity expression exists, where the approximation is also an index build-time parameter, affecting the overall performance~\cite{DBLP:journals/is/AumullerBF20}. 
Still, the per-tuple cost of probing the index is higher than scanning due to the data structure traversal and random accesses, aiming to be offset by lower overall cost. This assumption works when all the data is searched/joined. However, analytical queries are typically selective (on relational attributes), which is not directly compatible with a vector index. Rather, pre-filtering techniques are employed, where the result set excludes tuples based on the relational condition on the fly while still incurring the traversal cost~\cite{DBLP:conf/sigmod/WangYGJXLWGLXYY21}. On the other hand, a scan-based approach can exclude tuples at a lower cost.\end{RoneB}
\Rthree{We compare and evaluate our approach against the vector databases index approach (\autoref{sec:scan_or_probe}), extending the prior work on access path selection~\cite{DBLP:conf/sigmod/KesterAI17} and motivating for a selectivity-driven decision. We outline the key differences between the index and scan join in~\autoref{tab:comparison}.}


\subsection{Tensor Join Formulation: Enhancing The Nested-Loop Join}

\input{04-logical-optimization/tensor_formulation}

As the computation of context-enhanced operators happens over dense high-dimensional embedding vectors, following the vector and matrix definitions of cosine distance in \autoref{ssec:neur-emb}, we present the tensor formulation of the dot product. It is important to highlight that cosine similarity is equivalent to the dot product with normalized input vectors.

The tensor formulation allows reasoning about the potential decomposition of the problem for parallel and cache-efficient execution beyond data parallelism, a basis for the physical optimization (\autoref{sec:physical-optimization}). This enables efficient and well-studied matrix-based algorithms for linear algebra in addition to the traditional relational algorithms. We present the block-matrix decomposition of the problem~\cite{IMM2012-03274}. 

Given a $(|R| \times dim)$ matrix $\mathbf{R}$ with $t$ row partitions and $d$ column partitions, and a $(dim \times |S|)$ matrix $\mathbf{S}$ with $d$ row partitions and $v$ column partitions that are compatible with the partitions of $\mathbf{R}$, dot product $\mathbf{D} = \mathbf{R}\mathbf{S}$ can be formed block-wise, yielding $\mathbf{D}$ as a $(|R| \times |S|)$ matrix with $t$ row partitions and $v$ column partitions.
We consider $\mathbf{S}$ to be already transposed if the initial data layout is as of $\mathbf{R}$; in other words, matrices $\mathbf{R}$ and $\mathbf{S}$ are compatible.

In particular, we partition the data along the tuple lines, not the dimensions, as illustrated in \autoref{fig:tensor} \textcircled{1}. Transforming the initial Nested-Loop Join into Tensor formulation enables the application of linear algebra optimizations, in particular, matrix multiplication algorithms, to achieve better cache utilization of high-dimensional data with formalized parallelization using block-matrix decomposition. 
This is compatible with and extends recent research on formulating relational operators for tensor processing runtimes~\cite{DBLP:journals/pvldb/HeNBSSPCCKI22}, \Rone{as we explicitly consider linear algebra domain optimizations in our Tensor Join formulation}. 

In contrast to NLJ, a matrix block (several vectors) can remain in the cache and be reused over many operations, leading to better cache utilization. 
Block-matrix partitioning allows for defining the processing granularity, constraining the memory footprint, and allowing fine control of the processing granularity of cosine-distance-based similarity operations, all while reducing redundant data movement, resulting in a computationally and data access-optimized dense matrix operation.

The next step is to map back to corresponding tuple pairs that satisfy the \texttt{threshold} condition, as in \autoref{fig:tensor} \textcircled{2}. Maintaining the starting offsets of input relation partitions is sufficient, so the result set constitutes a potentially sparse matrix of pairs representing matrix batch offsets driven by the predicate selectivity.
This result can be considered equivalent to late materialization, and while sparse, it is more compact as tuples of offsets represent unique tensor identifiers. This is increasingly important when using novel memory hierarchies with fast but limited memory, such as high-bandwidth memory (HBM)~\cite{DBLP:conf/vldb/SancaA23}.\\

\textbf{Takeaway} Formulating the cost model and alternative equivalent execution plans using linear algebra allows tuning the algorithms to the cost model and execution environment parameters, as high-dimensional vectors and model processing introduce data access, caching, and processing overheads. This is a mandatory step that enables further logical and physical optimizations, different from the ones suitable for traditional relational operators that process only single-dimensional data.

%% file: 04-logical-optimization/tensor_formulation.tex
\begin{equation}
    \label{eq:decomposition}
    \tag{Block Matrix Dot Product Decomposition~\cite{IMM2012-03274}}
    \begin{aligned}
            \mathbf{D}_{tv} = \sum\limits_{i=1}^{d}\mathbf{R}_{ti} \mathbf{S}_{iv} \quad
            &
            \begin{bmatrix*}[r]
            \phantom{\;\;}\mathbf{S}_{11}& \dots & \phantom{\,}\mathbf{S}_{1v}\phantom{0}\\
            \phantom{00}\vdots \phantom{00}& \ddots & \vdots\phantom{00}\\
            \phantom{\;\;}\mathbf{S}_{d1}& \dots & \phantom{\,}\mathbf{S}_{dv}\phantom{0}\;  
            \end{bmatrix*}\rightarrow \mathbf{S} 
            \\
            \mathbf{R} \leftarrow 
            \begin{bmatrix*}[r]
            \mathbf{R}_{11}& \dots & \mathbf{R}_{1d} \\
            \vdots \phantom{0}& \ddots & \vdots\phantom{00}\\
            \mathbf{R}_{t1}& \dots & \mathbf{R}_{td}\;
            \end{bmatrix*}
            &
            \begin{bmatrix*}[r]
            \phantom{0}\mathbf{D}_{11}& \dots & \mathbf{D}_{1v}\phantom{0} \\
            \vdots\phantom{0} & \ddots & \vdots\phantom{00} \\
            \mathbf{D}_{t1} & \dots\ & \mathbf{D}_{tv}\phantom{0}
            \end{bmatrix*} \rightarrow \mathbf{D} = \mathbf{R}\mathbf{S}
    \end{aligned}
\end{equation}

%% file: 05-physical-optimization/main.tex
\label{sec:physical-optimization}
Modern data management systems are designed and optimized to efficiently utilize available hardware resources~\cite{DBLP:journals/pvldb/ChrysogelosKAA19, DBLP:conf/icde/ZukowskiWB12, DBLP:journals/pvldb/Neumann11}. Equally, machine learning and linear algebra frameworks are designed with physical optimizations to allow fast and efficient execution over vector data~\cite{abadi2016tensorflow, paszke2019pytorch, DBLP:conf/sigmod/WangYGJXLWGLXYY21, DBLP:journals/tbd/JohnsonDJ21}.

\subsection{Data-Parallel Execution}

To benefit from many core architectures, we outline the parallelization and hardware-conscious optimizations of the join algorithm. In contrast to the traditional Nested-Loop Join (NLJ) that allows exact cosine-distance-based joins, high-dimensional embedding vectors take up more space in the cache hierarchy. 
Consider a 32KB L1 cache, and we operate over 4-byte values. 
Using a 100-dimensional embedding vector, the L1 cache can store only 80 values, in contrast to 8000 for the single-dimensional data type. 
This necessitates cache-efficient implementation to benefit from the memory hierarchy.
Computing cosine distance over vectors requires more computation cycles than simply performing the regular value-based operation.
Thus, judicious use of hardware resources is necessary to speed up data access and computation.

\subsubsection{Data parallelization strategies}
Nested-Loop Join can be parallelized by partitioning the input relations and using a heuristic of keeping the smaller relation inside the inner loop to improve data and cache locality.
We propose using the matrix (tensor) formulation (\autoref{fig:tensor}) using linear algebra as an alternative to traditional NLJ.
In contrast to NLJ, matrix multiplication over dense vectors is a linear algebra operation with a better cache locality~\cite{DBLP:journals/toms/GotoG08, DBLP:conf/ipps/SmithGSHZ14}, improving the use of the memory hierarchy in the presence of high-dimensional data, and using efficient matrix multiplication algorithms and linear-algebra frameworks.

\subsubsection{CPU Hardware Support} Traditionally, CPUs benefit from the main memory access locality. They are general-purpose compute units designed to process full-precision data types (e.g., 32-bit and 64-bit) that can support SIMD, such as with Intel AVX instructions. Recent AVX-512~\cite{intelavx512fp16} instruction set has introduced hardware support for half-precision data types, which allows processing up to 32 16-bit floating point numbers in a SIMD register. 
\Rone{Accelerating machine learning operations is becoming more commonplace in modern and upcoming CPUs, such as 4th generation Intel Scalable Xeon Processors, which introduced specialized instruction sets (AMX) and registers meant for accelerating vector and matrix computations~\cite{nassif2022sapphire}, along with a limited capacity High-Bandwidth memory which can speed-up memory-bound access patterns~\cite{DBLP:conf/vldb/SancaA23}.}
In general, specialized instructions can accelerate the dense matrix computations. At the same time, low-latency access to memory enables optimizing the sparse matrix processing when processing the elements that satisfy the join predicate. 
As even the main memory is often limited or expensive, we propose how to constrain the memory requirements of the tensor join.

\subsubsection{SIMD Vectorization}
Executing linear algebra operations such as cosine distance over dense vectors is compute-intensive and involves repeated operations over every vector embedding element. 
Since operations such as sum are repeated over every element of the logical vector, it is a natural fit for using single-instruction multiple-data instructions (SIMD). 
We use SIMD vectorization supported by hardware to speed up the arithmetic operations using fewer processing cycles using specialized registers and compute units, in conjunction with data-parallel partitioning for multi-core operator execution. \Rthree{We also show that physical optimizations do not come for free and can be underutilized due to bad logical plans, emphasizing the additive performance proposed optimization steps.}



\subsection{Constraining the Memory Requirements} \label{sec:memory_requirements}

\begin{figure}
  \centering
  \includegraphics[width=0.55\linewidth]{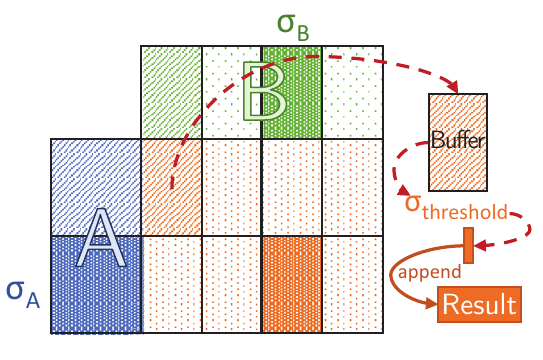}
  \caption{Matrix partitioning constrains the memory requirement.}
  \label{fig:filters_batched}
  \vspace{-15pt}
\end{figure}

The tensor join formulation (\cref{fig:tensor}) assumes a dot product operation between two matrices, a dense matrix linear algebra operation. This will result in a large intermediate state matrix of the same dimensions as the input relations. Despite joins being typically selective, which might reduce the matrix size, as in \cref{fig:filters_batched}, the intermediate state might still be too big to store and compute. Computing \texttt{|R|x|S|} for two 100k inputs yields a 100k x 100k matrix that results in 40GB of FP32 values. 
While this matrix can be preserved to offset future computation, even for modest input relation sizes, this approach, in its naive formulation, does not scale well concerning the memory requirements.

To resolve this issue, the previously presented matrix decomposition (\autoref{eq:decomposition}) enables partitioning the computation into batches and explicitly controlling the memory requirements based on the desired intermediate matrix size. This trades off memory for multiple invocations of the computation algorithm with smaller matrices, effectively computing the large one while pruning the intermediate sparse state after each sub-block-matrix computation.
We illustrate this in \autoref{fig:filters_batched}, where two relations \texttt{A} and \texttt{B} are joined over vectors. While the required memory requirement can be selectivity driven by other pushed-down relational predicates ($\sigma_A$, $\sigma_B$), this might not fit the available buffer budget. Thus, based on the available \textit{Buffer} size, the input data can be partitioned arbitrarily by decomposing the input data along the vector tuple boundaries (not dimensions). The strict \texttt{|A|x|B|} memory requirement becomes \texttt{Buffer = |part(A)|x|part(B)|}, at the cost of several invocations of the algorithm that might reduce the overall performance by frequent data movement and lower cache locality. 


\textbf{Takeaway} The physical operator design landscape encompasses implementation and hardware device characteristics-based decisions. 
Model-operator interactions only enrich and open a new design space.
High-dimensional data contributes to reduced capacity of the memory hierarchy in comparison to common atomic data types found in relational data processing and requires rethinking cache-local implementations.
With the increase in per-tuple compute cost, the strain is on both memory and compute resources, which invites the use of specialized hardware-conscious algorithms such as tensor join.

%% file: 06-evaluation/main.tex
\label{sec:evaluation}

We start by demonstrating the functionality of using models as a driver of context-enhanced relational operations through the example of word embeddings. 
We then focus on the main performance evaluation of the proposed logical and physical optimizations, showing that a holistic approach is necessary to obtain a performant join algorithm.

\textbf{System} To conduct the in-depth study, we implement our operators in a standalone system in C++ and use Intel AVX instructions for SIMD execution. Tensor formulation benchmarks use Intel oneAPI Math Kernel Library for CPU-aware and efficient BLAS-based linear algebra operations.
\Rone{We implement our index-based formulation in a hardware-optimized, open-source vector database Milvus~\cite{DBLP:conf/sigmod/WangYGJXLWGLXYY21}. We configure the system in standalone mode (version 2.3.9).}

\textbf{Hardware Setup} We run the end-to-end and scalability experiments on a two-socket Intel Xeon Gold 5118 CPU (2 x 12-core, 48 threads) and 384GB of RAM. All experiments are with in-memory data; experiments with synthetic data use the same random number generator seed for reproducibility. 

\subsection{Enhancing Operator Context via Word Embeddings}
In our study, we use the example of word embeddings that transform input strings into high-dimensional vectors. 
We show the context-awareness functionality that word embeddings allow and note that embedding models can be fine-tuned and replaced to support different notions of similarity. 
The intermediate data representation of an embedding is a context-free vector that operators process independently of the particular model, on top of which we base our analysis.
The proposed optimizations of our approach are independent by design and principled in approach due to the separation of concerns between the model, which produces vectors, and the operator performing the join over context-free vectors.

\textbf{Embedding Model}
We use FastText~\cite{DBLP:journals/tacl/BojanowskiGJM17, DBLP:conf/naacl/EdizelPBFGS19} as a model ($\mu$) for string embeddings, which has the desirable properties that it can be trained and adapted to the context, it supports out of vocabulary word embedding and is resilient to misspellings.  
\Rone{It learns word representations by extending Word2Vec~\cite{DBLP:journals/corr/abs-1301-3781} with subword information for better handling of rare words. It takes a string as an input and produces an embedding as an output.}
A context-aware operator is supplemented with an embedding model. In this case, when an operator receives strings, it embeds them using FastText and then performs the requested processing in the vector domain. 

\textbf{Dataset}
We train a 100-dimension embedding model over a subset of Wikipedia dataset~\cite{wiki}, cleaned of stopwords, using a subset of 1M strings from the dataset to test the similarity using the model.
We show the nearest vectors to sample words as the strings are embedded into a high-dimensional vector space. We then decode the vectors back into string and present sample semantic matches in \autoref{tab:sem_example}. The model has learned semantics and context from the Wikipedia dataset. To fine-tune the model, it is possible to specialize the embedding models with other domain-specific datasets.

\begin{table}[h]
  \caption{Semantic Matching using FastText trained on Wikipedia dataset, 100-D embeddings, sample words.}
  \label{tab:sem_example}
  \begin{tabularx}{\linewidth}{p{0.12\linewidth}  p{0.82\linewidth} }
    \toprule
    Word & Top-15 Model Matches\\
    \midrule
    dbms & rdbms, nosql, dbmss, postgresql, rdbmss, sql, dbmses, sqlite, dataflow, ordbms, oodbms, couchdb, mysql, ldap, oltp\\
    \hline
    postgres & postgre, postgresql, openvt, dbms, rdbmss, sqlite, dbmss, odbc, backend, rdbms, rdbmses, postgis, openvp, couchdb, mysql\\
    \hline
    clothes & dresses, clothing, garments, underwear, bedclothes, undergarments, towels, underwears, scarves, shoes, nightgowns, clothings, bathrobes, underclothes\\
  \bottomrule
\end{tabularx}
\end{table}

Models enable automated semantic matching, and the strings are not materialized or retrieved during operations in an intermediate step. The computation entirely happens on embedded data, and only positive matches are retrieved. It is possible to decode the embeddings, for example, based on their offset in the input relation and processing the embedding using standard encoder-decoder model architecture.

This model aimed to detect synonyms, semantic and related matches, and plural forms of the words without external user specification. The only parameter in the case of a join with cosine distance would be a single threshold parameter. This allows relational operators normally operating over sample strings (i.e., Word column in \autoref{tab:sem_example}) to perform matches with strings on the right in the embedding domain without humans in the loop or creating and specifying strict rules, \Rthree{in contrast to current similarity join approaches \cite{DBLP:journals/fcsc/YuLDF16,DBLP:journals/vldb/SilvaALPA13}}. Such models can work by providing positive match examples that could infer the correct cosine distance threshold parameter.


\subsection{NLJ Formulation: Logical Optimization}

\input{06-evaluation/cpu_noprefetch_endtoend} 

We extend the traditional relational join formulation by embedding vector processing and retrieval (\autoref{sec:logical-optimization}). 
We evaluate the impact of logical optimization of vector prefetching and physical optimization using SIMD in \autoref{fig:cpuNLJendtoendSIMD}. This experiment validates the cost difference between the naive join extension (\autoref{nlj-naive}) and the one aware of the vector retrieval (\autoref{nlj-prefetch}).
Not prefetching the embeddings incurs quadratic model access costs validating the cost model, resulting in orders of magnitude slower execution time \Rthree{and low benefit from hardware acceleration.}
This demonstrates the importance of analyzing, exposing, and optimizing model-operator interactions, where despite using the same hardware resources, \Rthree{including separate experiments with and without SIMD, the main bottleneck is not computational but access-pattern-related and algorithmic.} 
With the wrong holistic operator formulation, faster hardware cannot correct the suboptimal formulation, as may happen with imperative operator specification by a non-expert user.
In this case, the optimal strategy of prefetching the embeddings first and then joining, despite having two separate tasks, allows faster execution time. 
SIMD instructions improve the execution time 2x, indicating a computational bottleneck that additional hardware instructions reduce, while this is not possible in the non-prefetch, sub-optimal formulation.



\textbf{Takeaway.} Logical operator optimizations and task orchestration are crucial to removing algorithmic bottlenecks. Allocating more resources cannot scale and is wasteful before resolving an algorithm's logical costs and overheads. Using the improved NLJ cost model formulation, execution time scales linearly instead of quadratically, as in the non-optimized case.

\subsection{NLJ Formulation: Physical Optimization}

\input{06-evaluation/cpu_simd_scalability}

We focus next on the physical optimizations and demonstrate the scalability of CPU execution and physical and logical optimizations of NLJ formulation presented in \autoref{sec:physical-optimization}. 
First, we investigate the scalability (\autoref{fig:cpuScalabilitySIMD}). 
We enable hyperthreading (24 physical, 48 logical cores), affinitize threads to cores (2 threads will run on 1 physical core, 4 on 2, etc.), and run the NLJ formulation of 10k x 10k relation size input with 100-dimensional embeddings. 
The processor has AVX-512 registers that can simultaneously fit 16 32-bit floating-point values simultaneously. 
The average improvement is 5.36x, indicating non-computational overheads during vectorization but improved execution time using available hardware intrinsics.



\input{06-evaluation/cpu_endtoend}
Next, we evaluate the impact of different input relation sizes (in tuples) over 100-D, 32-bit embeddings over 48 threads and investigate the impact of physical and logical optimizations using the NLJ formulation.
In this experiment (\autoref{fig:cpuNLJendtoend}), we investigate the effects of input sizes, number of computations, and ordering of input relations of context-enhanced NLJ implementation. 
First, the execution time scales linearly with the number of computations/operations performed, according to the cost model (\autoref{nlj-prefetch}).
Second, we validate that to achieve improved execution time due to cache locality, a smaller relation should still be the inner loop, as in the traditional nested-loop-join. Despite more expensive per-vector computations, data access patterns still play an important role, impacting our experiment's performance by up to $\sim$35\% (at $10^{10}$ operations).

\textbf{Takeaway.} Logical and physical optimizations of the NLJ formulation with vectors enable reducing the overheads by orders of magnitude from the initial vector join extension. Still, the approaches we proposed until now optimize for vector execution without explicitly considering the high dimensionality and similarity operations over individual tuples. The tensor formulation, which we will evaluate next, addresses this issue.

\subsection{Tensor Formulation: The Holistic Vector-Join Optimization}

\input{06-evaluation/new_cpu_approaches}

We proposed batching multiple vector tuples in a tensor join formulation using optimized matrix computation (\autoref{fig:tensor}) instead of individual vector operations in the NLJ. 
The key enabler and difference is that resulting matrix operations are highly optimized for the cache locality that the simple NLJ imposed in its formulation. 
We evaluate this physical optimization proposed in \autoref{sec:physical-optimization}, evaluating whether the tensor formulation improves the per-vector-element processing time. 
We compare two strategies, running the fully optimized NLJ against the Tensor formulation. 
For this, we vary two factors: the total number of floating point numbers processed (\#FP32 Ops) and how many floating point numbers represent an individual vector (vector dimensionality, Vector \#FP32).
\autoref{fig:micro_approaches} summarizes the findings, where three data clusters are based on the number of operations, refined by individual vector size. 
In other words, for the 25600 case with dimensionality 1, there are $25600/1$ tuples joined, equally balanced in two relations, indicating $\sqrt{25600/1}=160$ tuples per input relation. Similarly, to obtain the number of tuples for the case of dimensionality 256, $\sqrt{25600/256}=10$. 
We use the per-FP32 breakdown as a unifying metric across the input size and dimensionality.
First, we notice the benefit of vectorization with increased vector size, where specialized hardware operations improve the per-tuple performance.
Second, pushing this boundary beyond per-tuple-vector but to a whole tuple-vector-batch (Tensor), when sufficient computation can benefit from the cache locality, significantly improves the execution time. The Tensor approach is slower only in case there were a few ($\sqrt{25600/64}=20$ and $\sqrt{25600/256}=10$) tuples to join.

\input{06-evaluation/new_cpu_gemm_batch}
Batching the vectors together in the tensor formulation is the key to reducing unnecessary data movement. 
We demonstrate the impact of batching in \autoref{fig:micro_batch}, where the BLAS-matrix operations are used with one fully batched relation. At the same time, the other is loaded vector-by-vector, repeated as many times as there are tuples. An alternative is where both relations are fully batched.
While inefficiencies are not noticeable with very small input sizes, batching becomes increasingly significant for scalability as the input grows.

\input{06-evaluation/new_memory_batch100k}

As explained in \autoref{sec:memory_requirements}, batching too many vectors in large tensors simultaneously comes at a prohibitive memory cost. 
We propose using mini-batches partitioned across tuple boundaries (\autoref{fig:filters_batched}) that can still benefit from the improved linear algebra algorithms and data locality.
The impact of batching is presented in \autoref{fig:100k_batching}.
We run the tensor join formulation over 100k x 100k, 100-D input using 48 threads. The \texttt{No Batch} case runs the join on the whole input simultaneously. 
At the same time, the experiment focuses on memory footprint reduction and the computational price to pay when various mini-batches are used.
While there is a negligible relative slowdown due to some added data movement and repeated operations, there is a significant benefit due to the reduction of the necessary memory.

\input{06-evaluation/new_tensor_nlj}
Finally, we compare the NLJ with the Tensor formulation end-to-end execution time in \autoref{fig:cpuTensor_NLJendtoend}.
While the execution time of both algorithms scales approximately linearly when increasing the input relation size, the algorithm optimizations enabled by batching vectors into tensors opened linear algebra-based execution optimizations with almost an order of magnitude improvement across various input sizes.

\begin{RoneB}
\subsection{Scan vs Probe: Vector Indexes Meet Analytical Workloads}
\label{sec:scan_or_probe}

With an optimized scan-based Tensor join, fully amenable to relational filtering, we compare this approach to a join operator implemented using vector indexes commonly used by vector databases that support relational pre-filtering. We construct an HNSW index~\cite{10.1109/TPAMI.2018.2889473}, taking the overall best-performing index from ANN-Benchmark~\cite{DBLP:journals/is/AumullerBF20}. As indexes are approximate and support the defined build-time distance, we construct the indexes meant for cosine-distance filtering with two cases: higher-recall/accuracy \textit{(Hi)} and lower-recall \textit{(Lo)}. \textit{Hi} has a maximum degree of nodes on each graph layer $M=64$, and search range parameter $efConstruction=512$. \textit{Lo} has $M=32$ and $efConstruction=256$, improving latency but reducing accuracy. The system enables parallel probing, which we use to implement a join by batching the search vector queries. It is mandatory to specify the top-k values to retrieve in an index-based approach, limiting the flexibility of the join operator. We limit the concurrent index probing to $10k$, construct an index on $1M$ vectors, and include one relational attribute column based on which we control the selectivity. We compare the scan-based tensor join with pre-filtering in the same setup, in-memory, and using 48 threads.

\input{06-evaluation/scan_vs_probe_top_k_1}
\input{06-evaluation/scan_vs_probe_top_k_32}

In case the index retrieves and joins only with the most similar value (\autoref{fig:scan_vs_probe_top_k1}), which is the best case for an index-based approach, the reduced scan and computation cost pays off at $20\%-30\%$ selectivity, where below this threshold the scan-based approach manages to effectively filter out and compute the join faster. Once we join with more than top-1 similar tuples, in this case, top-32 (\autoref{fig:scan_vs_probe_top_k32}), the index probe and traversal cost becomes more expensive, shifting the cross-point to $80\%$ for lower accuracy index, and impractical by being always slower for high-accuracy index. The scan-based approach is always exact, performing exhaustive comparisons.

The indexes are limited to build-time characteristics, and when a different expression, in this case (\autoref{fig:scan_vs_probe_range}), similarity coefficient filter ($similarity>0.9$) is applied, the index-based performance drops, being comparable to tensor join only around $5-10\% selectivity$ despite still retrieving tuples based on the top-k mechanism ($k=32$). On the other hand, Tensor-join is flexible concerning expression processing and returns not only top-k tuples in comparison but also all the matching and qualifying tuples, overall being faster in low and high selectivity ranges.

\input{06-evaluation/scan_vs_probe_range}


\end{RoneB}

\textbf{Takeaway.} Holistic optimization of the join algorithm with vector inputs is necessary to enable fast and efficient computation. We reduced model-operator overheads, tuned the individual and batched vector computation, and designed and evaluated access pattern-aware operators that efficiently use underlying hardware capabilities. \Rthree{We discussed access path selection~\cite{DBLP:conf/sigmod/KesterAI17} in vector data management and demonstrated the trade-offs in scan versus index-based join implementation.}

%% file: 06-evaluation/cpu_noprefetch_endtoend.tex
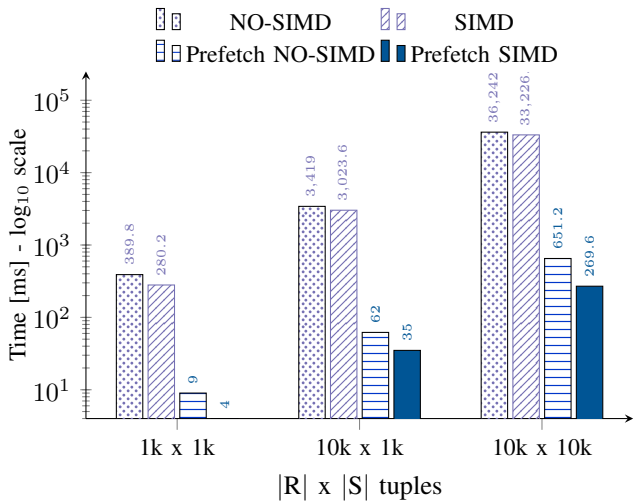
\begin{figure}
\centering
        \begin{tikzpicture}

        \begin{axis}[
            legend style={at={(0.5,1.1)},draw=none,
                anchor=south,
                legend columns=2,
                inner sep=0pt, outer sep=0pt, font=\small
            },
            axis lines = left,
            axis x line*=bottom,
            axis y line*=left,
            xlabel={$|$R$|$ x $|$S$|$ tuples},
            xlabel style={align=center},
            ylabel style={align=center},
            ylabel={Time [ms] - log$_{10}$ scale},
            ylabel absolute, every axis y label/.append style={yshift=-1em, font=\small},
            enlarge x limits=0.25,
            height=0.5\columnwidth,
            width=\linewidth,
            ymode=log,
            log basis y={10},
            ymin=0,
            xmin=1,
            xtick  = {1,2,3},
            xticklabels={
                1k x 1k, 
                10k x 1k,
                10k x 10k
            },
            bar width=10pt,
            x tick label style={font=\small,align=center},
            point meta=rawy,
            ymax=250000,
            ybar,
            nodes near coords,
            every node near coord/.append style={font=\tiny, rotate=90, anchor=west} 
        ]
        \addplot[fill=myPurple, text=myPurple, fill opacity=1, pattern=crosshatch dots, pattern color=myPurple]
        	coordinates {(1,389.8) (2,3419) (3,36242.6)};
        
        \addplot[fill=myPurple, myPurple, fill opacity=1, pattern=north east lines, pattern color=myPurple]
        	coordinates {(1,280.2) (2,3023.6) (3,33226.6)};
        
        \addplot[fill=ACMDarkBlue, text=ACMDarkBlue, pattern=horizontal lines, pattern color=ACMDarkBlue]
        	coordinates {(1,9) (2,62) (3,651.2)};

        \addplot[fill=ACMDarkBlue, text=ACMDarkBlue, fill opacity=1]
        	coordinates {(1,4) (2,35) (3,269.6)};
        
        \legend{NO-SIMD, SIMD, Prefetch NO-SIMD, Prefetch SIMD}
        \end{axis}
        \end{tikzpicture}

\caption{The impact of logical and physical optimization on NLJ formulation. 100-D vectors, 48 threads.}
\label{fig:cpuNLJendtoendSIMD}
\vspace{-10pt}
\end{figure}

%% file: 06-evaluation/cpu_simd_scalability.tex
\begin{figure}
\centering
        \begin{tikzpicture}

        \begin{axis}[
            legend style={at={(0.5,1)},draw=none,
                anchor=south,
                legend columns=2,
                inner sep=0pt, outer sep=0pt, font=\small
            },
            axis lines = left,
            axis x line*=bottom,
            axis y line*=left,
            xlabel={\# of threads},
            xlabel style={align=center},
            ylabel style={align=center},
            ylabel={Time [s]},
            ylabel absolute, every axis y label/.append style={yshift=-1em, font=\small},
            height=0.4\columnwidth,
            width=\linewidth,
            ymin=0,
            ymax=20,
            xmin=1,
            xmax=48,
            xtick={1,4,8,12,16,20,24,28,32,40,48},
        ]
            \addplot[color=ACMDarkBlue, thick]  coordinates {
                (1, 3.4266) 
                (2, 3.2852) 
                (3, 2.7918) 
                (4, 2.0956) 
                (8, 1.2246) 
                (12, 0.8538) 
                (16, 0.6252) 
                (20, 0.4986) 
                (24, 0.420) 
                (28, 0.3756) 
                (32, 0.3316) 
                (40, 0.2726) 
                (48, 0.2502) 
            };
            \addlegendentry{SIMD}
    
        	\addplot[color=ACMDarkBlue, thick, dashed]  coordinates {
        		(1, 18.6784) 
                (2, 17.3178) 
                (3, 11.675) 
                (4, 9.7206) 
                (8, 5.6294) 
                (12, 4.4454) 
                (16, 3.5488) 
                (20, 2.8122) 
                (24, 2.389) 
                (28, 2.1218) 
                (32, 1.9164) 
                (40, 1.5398) 
                (48, 1.5896) 
        	};
        	\addlegendentry{NO-SIMD}
	\end{axis}
    \end{tikzpicture}

\caption{Optimized NLJ scalability with correct logical optimization, 10k x 10k join input relations, 100-D vectors.}
\label{fig:cpuScalabilitySIMD}
\vspace*{-10pt}
\end{figure}
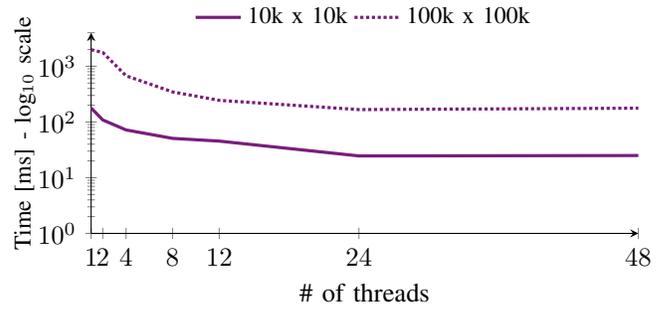

%% file: 06-evaluation/cpu_endtoend.tex
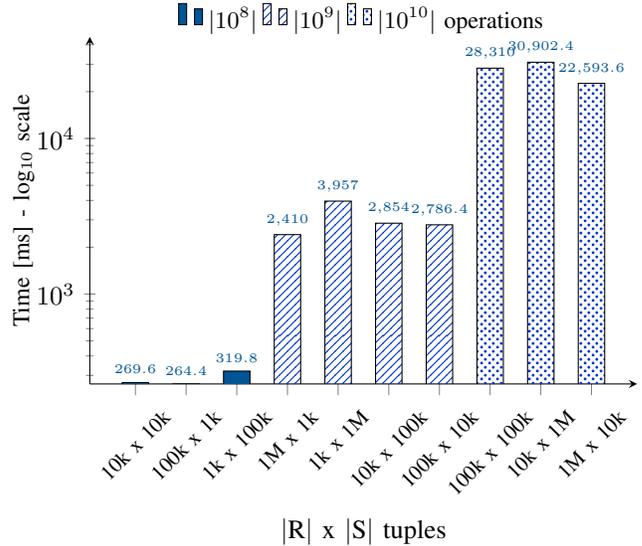
\begin{figure}
\centering
        \begin{tikzpicture}

        \begin{axis}[
            legend style={at={(0.5,1)},draw=none,
                anchor=south,
                legend columns=-1,
                inner sep=0pt, outer sep=0pt, font=\small
            },
            axis lines = left,
            axis x line*=bottom,
            axis y line*=left,
            xlabel={$|$R$|$ x $|$S$|$ tuples},
            xlabel style={align=center},
            ylabel style={align=center},
            ylabel={Time [ms] - log$_{10}$ scale},
            ylabel absolute, every axis y label/.append style={yshift=-1em, font=\small},
            enlarge x limits=0.1,
            height=0.6\columnwidth,
            width=\linewidth,
            ymode=log,
            log basis y={10},
            xmin=1,
            xtick  = {1,2,3,4,5,6,7,8,9,10 },
            xticklabels={
                10k x 10k, 
                100k x 1k,
                1k x 100k,
                1M x 1k,
                1k x 1M,
                10k x 100k,
                100k x 10k,
                100k x 100k,
                10k x 1M,
                1M x 10k
            },
            bar width=10pt,
            x tick label style={font=\footnotesize,rotate=45,align=center},
            point meta=rawy,
            ymax=45000,
            nodes near coords,
            every node near coord/.append style={font=\tiny},
            ybar
        ]
        \addplot[fill=ACMDarkBlue, text=ACMDarkBlue, fill opacity=1, xshift=+12pt]  
        	coordinates {(1,269.6) (2,264.4) (3,319.8)};
        
        \addplot[fill=ACMDarkBlue, text=ACMDarkBlue, fill opacity=1, pattern=north east lines, pattern color=ACMDarkBlue]
        	coordinates {(4,2410) (5,3957) (6,2854) (7,2786.4)};
        
        \addplot[fill=ACMDarkBlue, text=ACMDarkBlue, fill opacity=1, pattern=crosshatch dots, pattern color=ACMDarkBlue, xshift=-12pt]
        	coordinates {(8,28310) (9,30902.4) (10,22593.6)};
        
        \legend{$|10^8|$, $|10^9|$, $|10^{10}|$ operations}
        \end{axis}
        \end{tikzpicture}

\caption{Optimized NLJ formulation with varying input relation sizes, 100-D vectors, 48 threads.}
\label{fig:cpuNLJendtoend}
\vspace{-10pt}
\end{figure}

%% file: 06-evaluation/new_cpu_approaches.tex
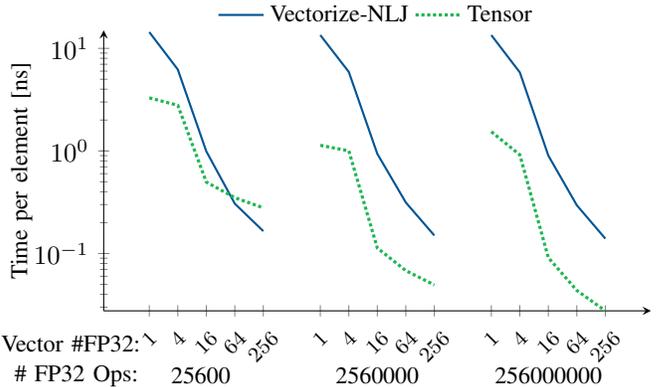
\begin{figure}
\centering

\begin{tikzpicture}
    \begin{axis}[
            legend style={at={(0.5,1)},draw=none,
                anchor=south,
                legend columns=-1,
                inner sep=0pt, outer sep=0pt, font=\small
            },
            axis lines = left,
            axis x line*=bottom,
            axis y line*=left,
            ylabel style={align=center},
            ylabel={Time per element [ns]},
            ylabel absolute, every axis y label/.append style={yshift=-0.4em, font=\footnotesize},
            enlarge x limits=0.1,
            height=0.4\columnwidth,
            width=\linewidth,
            ymode=log,
            log basis y={10},
            xmin=1,
            xtick={1,2,3,4,5,7,8,9,10,11,13,14,15,16,17},
            xticklabels={$1$,$4$,$16$,$64$,$256$,$1$,$4$,$16$,$64$,$256$,$1$,$4$,$16$,$64$,$256$},
            x tick label style={font=\footnotesize,rotate=45,align=center},
            point meta=rawy,
            ymax=15,
            clip=false
        ]

    \addplot+[color=ACMDarkBlue, thick, no markers] plot 
        coordinates {
            (1,14.4157552083333)
(2,6.2222265625)
(3,0.991145833333333)
(4,0.306171875)
(5,0.165221354166667)

(7,13.46841328125)
(8,5.89067122395833)
(9,0.937294140625)
(10,0.314390364583333)
(11,0.149983984375)

(13,13.4598139231771)
(14,5.83242371484375)
(15,0.90085184375)
(16,0.2952685625)
(17,0.13974198828125)
        };




    \addplot+[color=myGreen, very thick, densely dotted, no markers] plot
        coordinates {
            (1,3.294765625)
(2,2.78424479166667)
(3,0.497578125)
(4,0.350130208333333)
(5,0.2788671875)

(7,1.13450572916667)
(8,1.002325390625)
(9,0.113187630208333)
(10,0.0681109375)
(11,0.049408984375)

(13,1.54064754427083)
(14,0.916303227864583)
(15,0.0911605911458333)
(16,0.0436186510416667)
(17,0.0276135989583333)
        };
    
    \node[font=\small] at (xticklabel cs:.1775,0pt) {25600};
    \node[font=\small] at (xticklabel cs:.5000,0pt) {2560000};
    \node[font=\small] at (xticklabel cs:.8125,0pt) {256000000};
    
    \node[font=\small] at (xticklabel cs:-0.05,0pt) {\# FP32 Ops:};
    \node[font=\small] at (xticklabel cs:-0.06,-12pt) {Vector \#FP32:};

    \legend{Vectorize-NLJ, Tensor}
\end{axis}
\end{tikzpicture}

\caption{Physical optimization. The tensor strategy (green) pays off in larger inputs compared to NLJ (blue).}
\label{fig:micro_approaches}
\vspace{-10pt}
\end{figure}

%% file: 06-evaluation/new_cpu_gemm_batch.tex
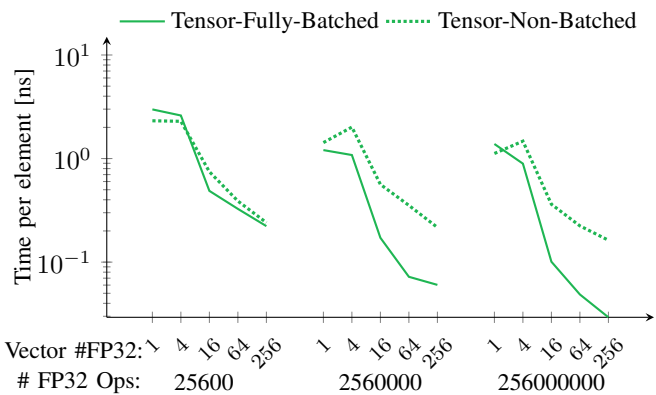
\begin{figure}
\centering

\begin{tikzpicture}
    \begin{axis}[
            legend style={at={(0.5,1)},draw=none,
                anchor=south,
                legend columns=-1,
                inner sep=0pt, outer sep=0pt, font=\small
            },
            axis lines = left,
            axis x line*=bottom,
            axis y line*=left,
            ylabel style={align=center},
            ylabel={Time per element [ns]},
            ylabel absolute, every axis y label/.append style={yshift=-0.4em, font=\footnotesize},
            enlarge x limits=0.1,
            height=0.4\columnwidth,
            width=\linewidth,
            ymode=log,
            log basis y={10},
            xmin=1,
            xtick={1,2,3,4,5,7,8,9,10,11,13,14,15,16,17},
            xticklabels={$1$,$4$,$16$,$64$,$256$,$1$,$4$,$16$,$64$,$256$,$1$,$4$,$16$,$64$,$256$},
            x tick label style={font=\footnotesize,rotate=45,align=center},
            point meta=rawy,
            ymax=15,
            clip=false
        ]

    \addplot+[color=myGreen, thick, no markers] plot
        coordinates {
(1,2.982265625)
(2,2.6049609375)
(3,0.4873828125)
(4,0.3266796875)
(5,0.222643229166667)

(7,1.20696471354167)
(8,1.081585546875)
(9,0.171814713541667)
(10,0.0721541666666667)
(11,0.0603)

(13,1.382621453125)
(14,0.893185283854167)
(15,0.100881559895833)
(16,0.0487078958333333)
(17,0.0293257669270833)
};

    \addplot+[color=myGreen, very thick, densely dotted, no markers] plot
        coordinates {
(1,2.31602864583333)
(2,2.2890234375)
(3,0.746432291666667)
(4,0.388815104166667)
(5,0.240794270833333)

(7,1.425690234375)
(8,2.01118098958333)
(9,0.562965885416667)
(10,0.353044010416667)
(11,0.216306901041667)

(13,1.11931083854167)
(14,1.47222652864583)
(15,0.362115166666667)
(16,0.223883802083333)
(17,0.1627293046875)
};
    
    \node[font=\small] at (xticklabel cs:.1775,0pt) {25600};
    \node[font=\small] at (xticklabel cs:.5000,0pt) {2560000};
    \node[font=\small] at (xticklabel cs:.8125,0pt) {256000000};
    
    \node[font=\small] at (xticklabel cs:-0.05,0pt) {\# FP32 Ops:};
    \node[font=\small] at (xticklabel cs:-0.06,-12pt) {Vector \#FP32:};

    \legend{Tensor-Fully-Batched, Tensor-Non-Batched}
\end{axis}
\end{tikzpicture}

\caption{The impact of vector batching. Non-batched indicates that one of the join inputs is processed one vector at a time.}
\label{fig:micro_batch}
\vspace{-15pt}
\end{figure}

%% file: 06-evaluation/new_memory_batch100k.tex
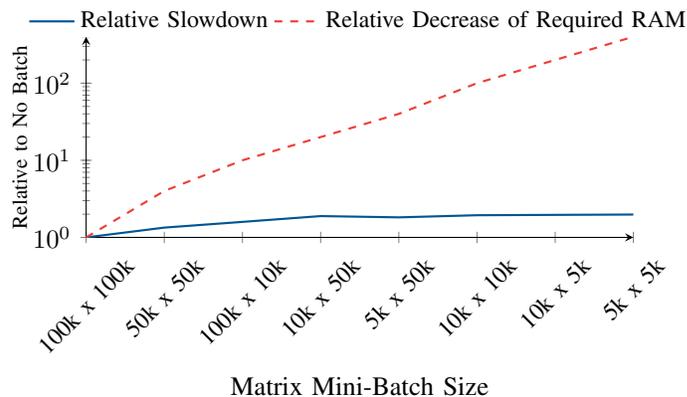
\begin{figure}
\centering
        \begin{tikzpicture}

        \begin{axis}[
            legend style={at={(0.5,1.1)},draw=none,
                anchor=south,
                legend columns=2,
                inner sep=0pt, outer sep=0pt, font=\small
            },
            axis lines = left,
            axis x line*=bottom,
            axis y line*=left,
            xlabel={Matrix Mini-Batch Size},
            xlabel style={align=center},
            ylabel style={align=center},
            ylabel={Relative to No Batch},
            ylabel absolute, every axis y label/.append style={yshift=-1em, font=\footnotesize},
            height=0.4\columnwidth,
            width=\linewidth,
            xmin=1,
            xmin=1,
            xtick  = {1,2,3,4,5,6,7,8,9},
            xticklabels={
                100k x 100k, 
                50k x 50k,
                100k x 10k,
                10k x 50k,
                5k x 50k,
                10k x 10k,
                10k x 5k,
                5k x 5k,
            },
            x tick label style={font=\small,rotate=45,align=center},
            ymode=log,
            log basis y={10},
        ]
            \addplot[color=ACMDarkBlue, thick]  coordinates {
                (1, 1) 
                (2, 1.34) 
                (3, 1.59) 
                (4, 1.89) 
                (5, 1.82) 
                (6, 1.94) 
                (7, 1.96) 
                (8, 1.98) 
            };
            \addlegendentry{Relative Slowdown}
    
        	\addplot[color=ACMRed, thick, dashed]  coordinates {
        		(1, 1) 
                (2, 4) 
                (3, 10) 
                (4, 20) 
                (5, 40) 
                (6, 100) 
                (7, 200) 
                (8, 400) 
        	};
        	\addlegendentry{Relative Decrease of Required RAM}
	\end{axis}
    \end{tikzpicture}

\caption{Batch size impact on memory requirements and execution time. 100k x 100k, 100-D input (No Batch case).}
\label{fig:100k_batching}
\vspace*{-10pt}
\end{figure}

%% file: 06-evaluation/new_tensor_nlj.tex
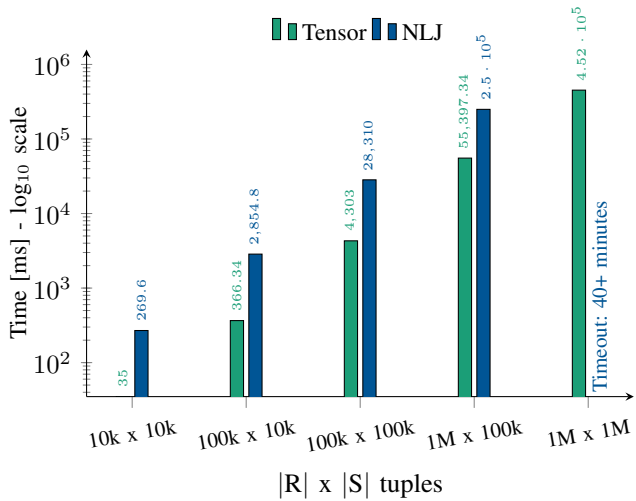
\begin{figure}
\centering
        \begin{tikzpicture}

        \begin{axis}[
            legend style={at={(0.5,1)},draw=none,
                anchor=south,
                legend columns=-1,
                inner sep=0pt, outer sep=0pt, font=\small
            },
            axis lines = left,
            axis x line*=bottom,
            axis y line*=left,
            xlabel={$|$R$|$ x $|$S$|$ tuples},
            xlabel style={align=center},
            ylabel style={align=center},
            ylabel={Time [ms] - log$_{10}$ scale},
            ylabel absolute, every axis y label/.append style={yshift=-1em, font=\small},
            enlarge x limits=0.1,
            height=0.55\columnwidth,
            width=\linewidth,
            ymode=log,
            log basis y={10},
            ymin=0,
            xmin=1,
            xtick  = {1,2,3,4,5},
            xticklabels={
                10k x 10k, 
                100k x 10k,
                100k x 100k,
                1M x 100k,
                1M x 1M
            },
            bar width=5pt,
            x tick label style={font=\footnotesize,rotate=10,align=center},
            point meta=rawy,
            ymax=1585000,
            ybar,
            nodes near coords,
            every node near coord/.append style={font=\tiny, rotate=90, anchor=west} 
        ]        
        \addplot[fill=myDarkGreen, text=myDarkGreen, fill opacity=1]
        	coordinates {(1,35) (2,366.34) (3,4303) (4,55397.34) (5,451837.67)};

        \addplot[fill=ACMDarkBlue, text=ACMDarkBlue, fill opacity=1]
        	coordinates {(1,269.6) (2,2854.8) (3,28310) (4,249673.6)};

        \node[rotate=90, anchor=west, text=ACMDarkBlue] at (220pt,30) {\footnotesize{Timeout: 40+ minutes}};
        
        \legend{Tensor, NLJ}
        \end{axis}
        \end{tikzpicture}

\caption{Tensor join vs. NLJ formulation, 100-D, 48 threads.}
\label{fig:cpuTensor_NLJendtoend}
\vspace{-15pt}
\end{figure}

%% file: 06-evaluation/scan_vs_probe_top_k_1.tex
\begin{figure}
\centering
        \begin{tikzpicture}

        \begin{axis}[
            legend style={at={(0.5,1)},draw=none,
                anchor=south,
                legend columns=2,
                inner sep=0pt, outer sep=0pt, font=\small
            },
            axis lines = left,
            axis x line*=bottom,
            axis y line*=left,
            xlabel={selectivity \%},
            xlabel style={align=center},
            ylabel style={align=center},
            ylabel={Time [s]},
            ylabel absolute, every axis y label/.append style={yshift=-1em, font=\small},
            height=0.4\columnwidth,
            width=\linewidth,
            ymin=0,
            xmin=0.01,
            xmax=100,
            xtick={0,10,20,30,40,50,60,70,80,90,100},
        ]
            \addplot[color=myGreen, thick]  coordinates {
                (100,2.25325)
                (99,2.48825)
                (95,2.443)
                (90,2.231)
                (80,2.13275)
                (70,1.868)
                (60,1.6655)
                (50,1.8055)
                (40,1.60375)
                (30.,1.29025)
                (20.,0.82625)
                (10.,0.434)
                (5.,0.21925)
                (1,0.0475)
            };
            \addlegendentry{Tensor Join}

            \addplot[color=myGreen, thick, dotted]  coordinates {
                (100,2.25325)
                (99,2.11525)
                (95,2.111)
                (90,1.873)
                (80,1.84875)
                (70,1.622)
                (60,1.4485)
                (50,1.6125)
                (40,1.43275)
                (30.,1.17625)
                (20.,0.73625)
                (10.,0.385)
                (5.,0.18225)
                (1,0.0435)
            };
            \addlegendentry{Tensor Join (-filter cost)}

            \addplot[color=ACMDarkBlue, thick, densely dotted]  coordinates {
                (100,0.725804138183593)
                (99,0.688578701019287)
                (95,0.685643339157104)
                (90,0.687362718582153)
                (80,0.673015642166137)
                (70,0.660204410552978)
                (60,0.659961605072021)
                (50,0.65615849494934)
                (40,0.657292175292968)
                (30.,0.651034498214721)
                (20.,0.702236986160278)
                (10.,0.811469602584838)
                (1,1.04758963584899)
        	};
        	\addlegendentry{Index Join (Lo)}

            \addplot[color=ACMDarkBlue, thick, dashed]  coordinates {
        		(100,1.76110572814941)
                (99,1.70281414985656)
                (95,1.70052714347839)
                (90,1.65744614601135)
                (80,1.564231300354)
                (70,1.47650299072265)
                (60,1.39712200164794)
                (50,1.29519710540771)
                (40,1.2073989868164)
                (30.,1.14377388954162)
                (20.,1.1922089099884)
                (10.,1.3853401184082)
                (1,1.04292845726013)
        	};
        	\addlegendentry{Index Join (Hi)}
	\end{axis}
    \end{tikzpicture}

\caption{Top-K=1 vector join condition (10k x 1M with filter)}
\label{fig:scan_vs_probe_top_k1}
\vspace*{-15pt}
\end{figure}

%% file: 06-evaluation/scan_vs_probe_top_k_32.tex
\begin{figure}
\centering
        \begin{tikzpicture}

        \begin{axis}[
            legend style={at={(0.5,1)},draw=none,
                anchor=south,
                legend columns=2,
                inner sep=0pt, outer sep=0pt, font=\small
            },
            axis lines = left,
            axis x line*=bottom,
            axis y line*=left,
            xlabel={selectivity \%},
            xlabel style={align=center},
            ylabel style={align=center},
            ylabel={Time [s]},
            ylabel absolute, every axis y label/.append style={yshift=-1em, font=\small},
            height=0.4\columnwidth,
            width=\linewidth,
            ymin=0,
            xmin=0.01,
            xmax=100,
            xtick={0,10,20,30,40,50,60,70,80,90,100},
        ]
            \addplot[color=myGreen, thick]  coordinates {
                (100,2.242)
                (99,2.412)
                (95,2.3155)
                (90,2.18825)
                (80,2.133)
                (70,1.7245)
                (60,1.62625)
                (50,1.567)
                (40,1.696)
                (30.,1.293)
                (20.,0.876)
                (10.,0.4165)
                (5.,0.22175)
                (1,0.04525)
            };
            \addlegendentry{Tensor Join}

            \addplot[color=myGreen, thick, dotted]  coordinates {
                (100,2.242)
                (99,2.039)
                (95,1.9835)
                (90,1.83025)
                (80,1.849)
                (70,1.4785)
                (60,1.40925)
                (50,1.374)
                (40,1.525)
                (30.,1.179)
                (20.,0.786)
                (10.,0.3675)
                (5.,0.18475)
                (1,0.04125)
            };
            \addlegendentry{Tensor Join (-filter cost)}
    
        	\addplot[color=ACMDarkBlue, thick, densely dotted]  coordinates {
                (100,2.1579879283905)
                (99,2.05445365905761)
                (95,2.128937625885)
                (90,2.07446122169494)
                (80,2.13937077522277)
                (70,2.04425935745239)
                (60,2.1159610748291)
                (50,2.08915958404541)
                (40,2.12003908157348)
                (30.,2.105633020401)
                (20.,2.19867014884948)
                (10.,2.31487383842468)
                (1,2.4316270828247)
        	};
        	\addlegendentry{Index Join (Lo)}

            \addplot[color=ACMDarkBlue, thick, dashed]  coordinates {
        		(100,3.34328727722168)
                (99,3.11458382606506)
                (95,3.14062070846557)
                (90,3.06376748085021)
                (80,3.01363840103149)
                (70,2.9072166442871)
                (60,2.89247074127197)
                (50,2.77150030136108)
                (40,2.77246723175048)
                (30.,2.66591668128967)
                (20.,2.68378748893737)
                (10.,2.88976626396179)
                (1,2.45498075485229)
        	};
        	\addlegendentry{Index Join (Hi)}

	\end{axis}
    \end{tikzpicture}

\caption{Top-K=32 vector join condition (10k x 1M with filter)}
\label{fig:scan_vs_probe_top_k32}
\vspace*{-10pt}
\end{figure}

%% file: 06-evaluation/scan_vs_probe_range.tex
\begin{figure}
\centering
        \begin{tikzpicture}

        \begin{axis}[
            legend style={at={(0.5,1)},draw=none,
                anchor=south,
                legend columns=3,
                inner sep=0pt, outer sep=0pt, font=\small
            },
            axis lines = left,
            axis x line*=bottom,
            axis y line*=left,
            xlabel={selectivity \%},
            xlabel style={align=center},
            ylabel style={align=center},
            ylabel={Time [s]},
            ylabel absolute, every axis y label/.append style={yshift=-1em, font=\small},
            height=0.4\columnwidth,
            width=\linewidth,
            ymin=0,
            xmin=0.01,
            xmax=100,
            xtick={0,10,20,30,40,50,60,70,80,90,100},
        ]
            \addplot[color=myGreen, thick]  coordinates {
                (100,2.32675)
                (99,2.74925)
                (95,2.60265)
                (90,2.29275)
                (80,2.00625)
                (70,2.051)
                (60,1.53825)
                (50,1.48375)
                (40,1.482)
                (30.,1.269)
                (20.,0.8845)
                (10.,0.43025)
                (5.,0.2535)
                (1,0.047)
            };
            \addlegendentry{Tensor Join}
    
        	\addplot[color=ACMDarkBlue, thick, densely dotted]  coordinates {
                (100,9.63729448318481)
                (99,9.39821147918701)
                (95,8.78228640556335)
                (90,7.94208669662475)
                (80,6.32952742576599)
                (70,5.11057653427124)
                (60,4.02779231071472)
                (50,2.92738437652587)
                (40,2.144176197052)
                (30.,1.44465508460998)
                (20.,0.914827108383178)
                (10.,0.506176710128784)
                (5.,0.377593564987182)
                (1,1.15174326896667)
        	};
        	\addlegendentry{Index Join (Lo)}

            \addplot[color=ACMDarkBlue, thick, dashed]  coordinates {
        		(100,19.0821241378784)
                (99,18.8949162960052)
                (95,17.8862218379974)
                (90,15.7562906742095)
                (80,13.4458126544952)
                (70,11.1149637699127)
                (60,8.55395617485046)
                (50,6.30253734588623)
                (40,4.44344010353088)
                (30.,2.88729214668273)
                (20.,1.75135264396667)
                (10.,0.892429256439209)
                (5.,0.504145288467407)
                (1,1.15095634460449)
        	};
        	\addlegendentry{Index Join (Hi)}
	\end{axis}
    \end{tikzpicture}

\caption{Range vector join condition (10k x 1M with filter)}
\label{fig:scan_vs_probe_range}
\vspace*{-15pt}
\end{figure}

%% file: 07-related-work/main.tex
\label{sec:related-work}
This section outlines the related work and compares and places our approach in the rich design space of prior research.

\textbf{Machine Learning for Databases} has been a research topic where DBMS's structural components get enhanced using ML community findings. 
Learned indexes~\cite{DBLP:conf/sigmod/KraskaBCDP18} avoid data structure traversal by learning the data distribution information and optimizing data access. 
From a systems perspective, using tensor processing frameworks for traditional relational processing has also recently been proposed~\cite{DBLP:journals/corr/abs-2211-02753}.
Our approach is similar as we propose using ML embedding models to provide context to data traditionally opaque to relational DBMS, \Rone{while broadening the analysis and application scope of traditional analytics with novel model-relational interactions compatible with the relational model.}

\textbf{Databases for Machine Learning}
focus on applying or integrating machine learning components with systems. 
Frameworks such as Tensorflow~\cite{abadi2016tensorflow} or Pytorch~\cite{paszke2019pytorch} are efficient, hardware-conscious dataflow engines.
\Rone{The proliferation of model-embeddings fueled the development of vector-specialized databases~\cite{DBLP:conf/sigmod/WangYGJXLWGLXYY21}, focusing on retrieval and search with efficient high-dimensional similarity indexes~\cite{DBLP:journals/tbd/JohnsonDJ21}.}
Still, such engines often lack a DBMS's expressiveness, functionality, and analytical operations for more complex data analysis, which would force users to write imperative code to integrate different siloed system components, \Rone{with suboptimal relational filtering and access path selection.}
\Rone{While systems that combine analytics with training and deploying machine learning models exist~\cite{GoogleCloud2023BigQueryML,MicrosoftAzureML}, to our best knowledge, they only provide usability wrappers around models and limited User-Defined Function (UDF) capabilities~\cite{GoogleCloud2024BigQueryUDF} for implementing custom operators, making them an impractical platform for the end-to-end optimizations and detailed sensitivity analysis.}


\Rtwo{\textbf{Similarity Joins.}}
We proposed a join operator that is functionally a similarity join, \Rtwo{as processing embeddings inherently entails similarity distance computation instead of comparing exact values}.
Similarly, locality-sensitive hashing techniques~\cite{10.1145/3292500.3330853} exist for approximate joins. Traditional string similarity techniques require exact similarity specification using edit distance or q-grams as token-based string similarity~\cite{10.5555/645927.672200}, requiring users to specify the similarity rules for finding misspellings or token-based differences~\cite{DBLP:journals/vldb/SilvaALPA13,DBLP:journals/fcsc/YuLDF16}.

In contrast, we propose using word embedding models capable of identifying misspellings, different tenses, and semantic similarity based on training and fine-tuning training dataset and parameters~\cite{DBLP:journals/tacl/BojanowskiGJM17, DBLP:conf/naacl/EdizelPBFGS19}. Through the separation of concerns, from the perspective of DBMS, string similarity join has a tensor-based input with cosine distance and threshold as parameters. At the same time, the embedding model handles the string semantics and context and transforms the input into context-free embeddings for RDBMS to process.
Furthermore, our approach extends the notion of similarity to other context-rich data formats and modalities for which embedding models exist~\cite{DBLP:journals/jmlr/RaffelSRLNMZLL20,DBLP:journals/taslp/KongCIWWP20,DBLP:conf/naacl/DevlinCLT19,DBLP:conf/naacl/EdizelPBFGS19,DBLP:journals/corr/abs-1301-3781,DBLP:conf/cvpr/HeZRS16,DBLP:journals/corr/abs-2108-07258}.

\textbf{Representation Learning}
The significant body of work in representation learning is the key enabler of context-rich relational operators. 
It allows for transforming the human-centric, context-rich data representations into machine-centric formats amenable to automated processing.
We combine ML-based embedding models with relational operators and analyze end-to-end interactions, from logical to physical optimizations. Such models allow masking and transforming contextual data into embeddings as ubiquitous data representations that can be processed by extending RDBMS with tensor-based operators.
A rich research area in machine learning drives embedding models that support other context-rich data formats beyond strings, equally transforming the input into context-free embeddings, enabling multi-modality~\cite{DBLP:journals/jmlr/RaffelSRLNMZLL20}, processing images~\cite{DBLP:conf/cvpr/HeZRS16}, audio~\cite{DBLP:journals/taslp/KongCIWWP20}, documents~\cite{DBLP:conf/naacl/DevlinCLT19,DBLP:conf/rep4nlp/ChenSPM17}. Models trained on web-scale data exist as Foundation Models~\cite{DBLP:journals/corr/abs-2108-07258}, that can be re-trained and adapted for a specific task and dataset.

%% file: 08-conclusion/main.tex
\label{sec:conclusion}
Data management systems support analysts with modern data processing tools. 
As embedding models automate context-rich analysis, RDBMS needs to provide operators based on extended relational algebra and optimize their interactions.
Operations over the embeddings are not exact but intrinsically fuzzy and necessitate similarity operations. We propose a context-enhanced join based on the key observation of the separation of concerns between the context-providing embedding model and the analytical engine. 
We analyze the behavior of the join operator and propose relational algebra extensions, as well as logical and physical optimizations for efficient execution.
This includes the comparison of our scan-based Tensor Join against the vector index-based alternative in a hybrid vector-relational setting, demonstrating the impact of optimizations and access path selection trade-offs on the execution time. 
